\documentclass{aa}
\usepackage{ifthen}
\newcounter{refereemode}
\usepackage{graphicx}
\usepackage{amssymb}
\usepackage{natbib}
\bibpunct{(}{)}{;}{a}{}{,}
\usepackage{multirow}

\def\S{Sect.}

\newcommand{\Log}{\mbox{Log}}

\newcommand{\xmm}{XMM-{\em Newton}}
\newcommand{\chandra}{{\em Chandra}}

\newcommand{\lognlogs}{Log~$N$--Log~$S$ }
\newcommand{\lognlogsa}{Log~$N$--Log~$S$}

\newcommand{\LX}{L_{\rm X} }
\newcommand{\Lsx}{L_{0.5-2} }

\newcommand{\de}{{\rm d}}
\newcommand{\ergs}{erg s$^{-1}$}
\newcommand{\ergscmq}{erg s$^{-1}$ cm$^{-2}$}

\newcommand{\ergsHz}{erg s$^{-1}$ Hz$^{-1}$}
\newcommand{\nWmqsr}{nW m$^{-2}$ sr$^{-1}$}
\newcommand{\e}[1]{\cdot 10^{#1}}
\newcommand{\figuraacolori}{{\em (In colour only in the electronic edition)}\/\ }
\newcommand{\fxott}{\Log (F_{{\rm X-ray}}/F_{{\rm opt}})}
\newcommand{\etalum}{{\eta_{\rm l}}}
\newcommand{\etaden}{{\eta_{\rm d}}}

\begin{document}

\title{The X-ray luminosity function and number counts\\ of spiral galaxies}

\author{Piero Ranalli\inst{1,2} \and Andrea Comastri\inst{1} \and 
  Giancarlo Setti\inst{2,3} } 

\offprints{Piero Ranalli,\\ \email{piero.ranalli@bo.astro.it}}

\institute{
  INAF -- Osservatorio Astronomico di Bologna,
  via Ranzani 1, I--40127 Bologna, Italy
\and 
  Universit\`a di Bologna, Dipartimento di Astronomia, 
  via Ranzani 1, I--40127 Bologna, Italy
\and
  INAF -- Istituto di Radioastronomia,
  via Gobetti 1, I--40129 Bologna, Italy
}

\date{Received; accepted}

\abstract{ We discuss the X-ray luminosity function and number counts
  of normal spiral galaxies. A detailed comparison is performed of the
  LFs compiled at infrared, radio and optical wavelengths and
  converted into XLFs using available relations in the literature with
  the XLF directly estimated in the 0.5--2 keV energy band from X-ray
  surveys \citep{colin}. We find that the XLF from the local sample of
  IRAS galaxies \citep{takeu03} provides a good representation of all
  available data samples; pure luminosity evolution of the form
  $(1+z)^\eta$, with $\eta\lesssim 3$, is favoured over pure density
  evolution. We also find that the local X-ray luminosity density is
  well defined, $3\e{37} \pm 30\%$ \ergs\ Mpc$^{-3}$.
  
  
  We discuss different estimates of the galaxies \lognlogsa, selected
  from the \chandra\ Deep Field surveys with different selection
  criteria, and find that these have similar slopes, but
  normalisations scattered within a factor $\sim$ 2, of the same order
  of the Poissonian error on the number counts.  We then compare the
  observed galaxies \lognlogs with the counts predicted by the
  integration of our reference $z=0$ XLF. From the analysis of number
  counts alone, it is not possible to discriminate between density and
  luminosity evolution; however, the evolution of galaxies must be
  stopped in both cases at a redshift $z\sim 1$--2. The
    contribution of galaxies to the cosmic X-ray background is found
    to be in the range $6\%$--$12\%$.  By means of a complementary
    analysis with cosmic star formation models, we also find that the
  observed X-ray number counts might be not compatible with very large
  star formation rates at $z\sim 3$ as suggested by sub-mm
  observations in \citet{blain99a}.
  
  Concerning the content of current and, possibly, future X-ray
  surveys, we determine the fraction of normal galaxies around the
  current flux limit. At $\sim 3\e{-17}$ \ergscmq\ in the 0.5--2.0 keV
  band, the normal galaxies in the \chandra\ Deep Field surveys are
  about the $(30\pm 12)\%$ of the total number of objects.  At fainter
  fluxes the fraction of galaxies will probably rise, and overcome the
  counts from AGN at fluxes $\lesssim 10^{-17}$ \ergscmq.


}

\ifthenelse{\value{refereemode}=1}{
\titlerunning{X-ray \lognlogs and luminosity function of galaxies}
}{}

\maketitle

\section{Introduction}
\label{sect_norman}

The X-ray luminosity of normal spiral galaxies (hereafter `galaxies')
appears to be a reliable, absorption-free estimator of star formation
(i.e.\ their X-ray luminosity is dominated by emission from star
formation related processes, and not by an AGN). Growing attention has
been paid to its relation with emission at other wavelengths usually
taken as star formation indicators, such as the radio and far infrared
(FIR) bands, in both the local and the distant universe; it has been
found (\citealt{shapley01,bauer02,rcs03}, hereafter RCS03) that the
X-ray, radio and FIR luminosities are linearly and tightly correlated.
Since the FIR and radio luminosities are usually taken as Star
Formation Rate (SFR) indicators \citep{kenn98,cond92}, it has been
proposed that the X-ray luminosity also is a SFR indicator
(RCS03). 

Thus, some authors \citep{nandra02,colin} have begun to explore the
X-ray emission as a tool to investigate the Cosmic Star Formation
History (CSFH). To this end, the study of the X-ray luminosity
function (XLF) of galaxies and of its evolution represents a necessary
step. While the number counts may not lead directly to a determination
of the CSFH, they are indeed useful as a testbed to constrain the
evolution of the luminosity function (LF) at redshifts larger than
those directly probed by the observations.


Some early attempts to derive an XLF \citep{schmidt1996,ioannis99}
were mainly limited by problems in selecting the galaxy samples.
Nowadays, the superb imaging capabilities of the ACIS detectors
onboard Chandra has allowed the detection of extremely faint X-ray
fluxes ($\sim 10^{-17}$ \ergscmq\ in the 0.5--2 keV band), thus
probing the X-ray emission from low luminosity AGN and galaxies at
moderate redshifts ($z\lesssim 1$).  However, in order to distinguish
between accretion- and star formation-powered systems, spectroscopic
and photometric follow-up observations are necessary.  While these
efforts are ongoing, the difficulty to unambiguously classify the
nature of the X-ray emission, makes a detailed determination difficult
of the XLF properties rather hard.

An attempt to overcome these difficulties, based on photometric
selection criteria, has been discussed in \citet*{colin}, where a
sample was defined containing 210 galaxies with known redshift from
the \chandra\ Deep Field catalogues \citep{alexander03,giacconi02}.  A
Bayesian approach was chosen to derive a selection probability from
the values of three different parameters:
\begin{itemize}
\item the 0.5--2.0 keV X-ray luminosity for star forming galaxies is usually
  less than $\sim 10^{42}$ \ergs;
\item star forming galaxies have a softer spectrum than AGN; this
  translates in selecting objects with an hardness ratio $HR\lesssim -0.8$
  (with $HR=(H-S)/(H+S)$, where $H$ and $S$ represent the 2.0--10 and
  0.5--2.0 keV fluxes, respectively);
\item an X-ray/optical (R band) logarithmic flux ratio $\fxott<-1$ (a
  Log flux ratio of $-1$ is usually taken as an approximate boundary
  between galaxies and Seyferts; \citealt{maccacaro88}).
\end{itemize}


Two redshift bins ($z\leq 0.5$ and $0.5<z\leq 1.2$ with mean redshifts
$\bar z=0.27$ and 0.79, respectively) were considered in order to have
a comparable number of galaxies in both bins. A binned LF was derived
with the method developed in \citet{pageca00}, which is a variant of
the classical $1/V_{\rm max}$ method by \citet{schmidt68}.  The
evolution of the XLF, parameterised with a single power law in both
the redshift bins, is adequately described by a pure luminosity
evolution of the form $(1+z)^\etalum$ with $\etalum\sim 2.7$ (error
not available). It should be noted, however, that for a power law LF
it is not possible to discriminate between density and luminosity
evolution.

A robust estimate of the XLF and of the cosmological evolution of star
forming galaxies would greatly benefit from a detailed investigation
of the relative merits of different selection criteria.  For example,
the Bayesian approach developed by \citet{colin} might be affected by
some contamination from low luminosity AGN with a soft X-ray spectrum.
Well defined criteria to select large samples of star forming galaxies
are available in the infrared and radio bands, but the lack of X-ray
observations prevents a direct determination of their XLF.
Nevertheless, it appears feasible to attempt an inter comparison of
the LFs obtained in the radio, mid/far infrared and optical bands for
statistically significant samples, and in a redshift range as wide as
possible, by transforming each of them in a corresponding XLF.  This
multi-wavelength approach builds on the well-known linear relations
between X-ray, radio and far infrared luminosities calibrated for a
well defined sample of objects, which we reported in RCS03, and also
on non-linear relations between blue and X-ray/radio/FIR luminosities
reported by \citet{shapley01}.


Thus the structure of this article is as follows. In
\S~\ref{sect_FIR2XLF} we review the current determinations of
infrared, radio, and optical (B band) LFs.  In
\S~\ref{sec:xlf-discussion}, the LFs are converted into the X-ray
band, and the predicted local ($z=0$) XLFs are discussed; the
predicted XLFs are also compared with the one observed by
\citet{colin}. In \S~\ref{sect_obslogn} we review the observed X-ray
\lognlogs of galaxies based on different selection criteria. We recall
from RCS03 the predicted X-ray counts based on the radio \lognlogs and
compare them with the observed X-ray counts. We also compare the
observed counts with those predicted by the integration of the LFs. In
\S~\ref{sec:csfh}, the \lognlogs allow us to derive constraints on the
LF evolution and on the cosmic star formation history. The
contribution of the considered population of galaxies to the cosmic
infrared background is also calculated as used to further constrain
the LF evolution.  Finally, in \S~\ref{sec:counts-finalremarks} we
discuss the implications for the XLF of star forming galaxies and for
its evolution based on the previous analysis, we derive the fraction
of galaxies in current deep X-ray survey and discuss the implications
for future surveys and missions.

If not otherwise indicated (e.g., Table~\ref{tab:LFs} where LF
parameters are given in an $H_0$-independent form), we assume
$H_0=70$ km s$^{-1}$ Mpc$^{-1}$, $\Omega_{\rm M}=0.3$ and
$\Omega_{\Lambda}=0.7$.



\section{The LFs of star forming galaxies}
\label{sect_FIR2XLF}

We will consider the local differential luminosity function
$\varphi(L) \de \Log L$, which is the comoving number density of
sources per logarithmic luminosity interval, subject to either a pure
density evolution of the form $\varphi(z)\propto (1+z)^{\eta_{\rm d}}$,
or a pure luminosity evolution of the form $L(z)\propto
(1+z)^{\eta_{\rm l}}$.


\subsection{The FIR LFs}

Infrared surveys provide a powerful method to select star forming
galaxies, since the bulk of the far and near infrared emission is due
to reprocessed light from star formation, with AGN representing only a
minor population \citep{dejong84,franceschini01,elbaz02}.  In the
following, we summarise the current determinations of the FIR
luminosity functions.

The FIR LFs discussed below can be safely assumed to be essentially
unaffected by a contribution from Seyfert galaxies. By considering the
$60\mu$ LF of Seyfert galaxies \citep{rush}, we estimate the fraction
of Seyferts in the IRAS and ISO LFs discussed below to be $\sim 5\%$
for galaxies with $\nu L_\nu\sim 10^{10} L\sun$, and $\sim 10\%$ for
galaxies with $\nu L_\nu\sim 10^{12} L\sun$.

\begin{figure}[tp]    
  \centering \includegraphics[width=\columnwidth]{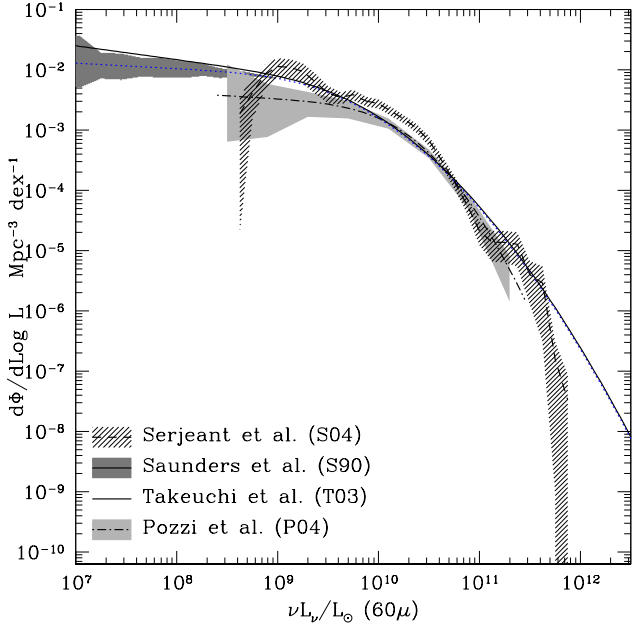}
  \caption { Infrared local luminosity functions.  Solid
  curve: IRAS 60$\mu$ LF from T03.  Dotted curve and dark grey area:
  IRAS 60$\mu$ LF from \citet{saun90}, with errors from $1/V_{\rm
  max}$ analysis. The two IRAS LFs coincide at bright
  luminosities. The size of the errors on the T03 LF and on the knee
  region of \citeauthor{saun90} is comparable with the line width.
  Dashed curve: ISO 90$\mu$ LF from S04; the shaded area corresponds
  to the 1$\sigma$ error bar. Dot-dashed curve: ISO 15$\mu$ LF from
  P04; the light grey area shows the $1\sigma$ error bar.  While the
  IRAS LFs are well constrained in the whole range of luminosities
  shown here, the ISO 90$\mu$ LF suffers from incompleteness for $\nu
  L_\nu\lesssim 10^9 L\sun$.  The monochromatic luminosities at $90\mu$
  have been converted to bolometric ones by assuming $\nu L_\nu = {\rm
  cost}$. The 15$\mu$ luminosities have been converted by assuming
  $L_{60\mu}/L_{15\mu}\sim 5$ \citep{mazzei01}.  $H_0=70$ is assumed.
  \label{fi:infratake} }
\end{figure}

\citet{saun90} defined a sample of 2,818 galaxies matching common
selection criteria from different IRAS samples, with a flux limit
around 0.6 Jy at 60$\mu$, complete at the 98\% level, redshifts
$z\lesssim 0.05$, and derived a $60\mu$ luminosity function. The
best-fit model has the shape
\begin{equation}
  \varphi(L)=\varphi^* \left( \frac{L}{L^*}
  \right)^{1-\alpha}
  \exp \left[ -\frac{1}{2\sigma^2} \Log^2_{10} \left( 1+\frac{L}{L^*}
      \right) \right] 
\label{eq:saundersfunct}\end{equation}
with the parameters reported in Table~\ref{tab:LFs}.
The evolution was parameterised as pure density evolution and the
exponent found to be
$\eta_{\rm d}=6.7\pm 2.3$.
The LF is shown as the dotted curve in Fig.~\ref{fi:infratake}, with
errors from $1/V_{\rm max}$ analysis as the dark grey area.


The $60\mu$ luminosity function was revised by
\citet[][hereafter T03]{takeu03,takeu03err}
by enlarging the galaxy sample: 15,411 galaxies from the Point Source
Catalog Redshift (PSC$z$, \citealt{saun00}) were used, covering 84\%
of the sky with a flux limit of 0.6 mJy at $60\mu$.
The same parameterisation of Eq.~(\ref{eq:saundersfunct}) was used for
the LF (Table~\ref{tab:LFs}).
In Fig.~\ref{fi:infratake} it is shown as the solid curve. This LF
coincides with the \citet{saun90} one for $\nu L_\nu\gtrsim 10^9L\sun$
and only differs by a factor of 2 at $10^7 L\sun$. It was also found
that the density evolution proposed by \citet{saun90} is not
consistent with this sample: a milder evolution, $\eta_{\rm d}=3.4\pm
0.70$, is reported to be the best-fit description of the data. Since
this represents an update of \citet{saun90} LF, in the following we
shall refer only to the T03 LF. Note that the redshift distribution of
the PSC$z$ galaxies (which is a superset of \citeauthor{saun90}
sample) spans a very limited range ($z\lesssim 0.07$), so that the
evolution exponent is poorly constrained.

The LF at $90\mu$ was recently determined by the ELAIS survey team
\citep[][hereafter S04]{serje01,serje04} with the PHOT instrument
onboard the ISO satellite: 151 objects detected in an area of 7.4
square degrees with flux limit 70 mJy and spectroscopic redshifts
(most of the objects having $z\lesssim 0.3$) were used to derive the
LF. It has not been fitted to a parametric form; see the dashed curve
in Fig.~(\ref{fi:infratake}) for the LF and Table~\ref{tab:LFs} for
the evolution.
To convert the ELAIS LF to $60\mu$, a $S_{60\mu}/S_{90\mu}=0.66$ flux
ratio can be derived by assuming $\nu L_\nu={\rm cost}$, which well
reproduces the spectrum of star forming galaxies in the wavelength
region here considered.


\begin{table*}[tp]
  \caption{Luminosity functions of galaxies at infrared,
    radio and blue wavelengths.}
  \label{tab:LFs}
  \centering
  \begin{tabular}{cllll}
\hline\hline
Band &Author &LF form    &Parameters    &Evolution\\
\hline
\multirow{2}*{$60\mu$} &Saunders
&\multirow{2}*{Eq.~(\ref{eq:saundersfunct})}
&$\varphi^*=(2.6\pm 0.8)\cdot 10^{-2}\, h^3\, {\rm Mpc}^{-3}, \alpha=1.09 \pm 0.12,$
&\multirow{2}*{$\etaden=6.7\pm2.3$}\\ 
&et al.
& &$ \sigma=0.72\pm 0.03 \mbox{ and } L^*=(2.95_{-1.21}^{+3.06})\cdot
10^{8}\, h^{-2}\, L\sun$\bigskip\\
\multirow{2}*{$60\mu$} &Takeuchi
&\multirow{2}*{Eq.~(\ref{eq:saundersfunct})}
&$\varphi^*=(2.34\pm 0.30)\cdot 10^{-2}\, h^3\, {\rm Mpc}^{-3}, \alpha=1.23 \pm 0.04,$
&\multirow{2}*{$\etaden=3.4\pm0.7$}\\
&et al.\ (T03)
& &$\sigma=0.724\pm 0.01\mbox{ and }L^*=(4.4\pm 0.9)\cdot 10^8\, h^{-2}\,
L\sun$ \bigskip\\
\multirow{2}*{$90\mu$} &Serjeant
&\multirow{2}*{numerical, see Fig.~(\ref{fi:infratake})\kern-4em}
& &\multirow{2}*{$\etaden=3.5\pm1.1$}\\
&et al.\ (S04) &  &\bigskip\\
\multirow{6}*{$15\mu$} &
& &quiescent spirals:   
&\multirow{2}*{\valign{#&#\cr spirals: no evolution\cr}}\\
& & &$\varphi^*=3.55\e{-3} \, h^3\, {\rm Mpc}^{-3}, \alpha=1.10\pm 0.25,$\\
&Pozzi &Eq.~(\ref{eq:saundersfunct}) spirals $+$ &$\sigma=0.5^{+0.1}_{-0.2}\mbox{ and }L^*=(6.3^{+25}_{-5.5})\cdot 10^8\, h^{-2}\, L\sun$
&starbursts: \\
&et al.\ (P04) &Eq.~(\ref{eq:saundersfunct}) starbursts &starbursts: & $\etaden=3.8\pm2$ \\
& & &$\varphi^*=2.95\e{-4}\, h^3\, {\rm Mpc}^{-3}, \alpha \mbox{ fixed at }0,$
&{\hspace{3em}\em with}\\
& & &$\sigma=0.39\pm 0.025\mbox{ and }L^*=(6.3^{+6.3}_{-2.3})\cdot 10^8\, h^{-2}\,L\sun$ 
& $\etalum={3.5_{-3.5}^{+1.0}}$ \bigskip\\
\multirow{3}*{1.4GHz} &Machalski \&\kern-1em
&\multirow{3}*{Eq.~(\ref{eq:saundersfunct})}
&$\varphi^*=(7.9_{-2.6}^{+3.8})\e{-3}\, h^3\ {\rm Mpc}^{-3} {\rm dex}^{-1},$
&constraints are too loose\\
&Godlowski & &$ \alpha=1.22 \pm 0.27, \sigma=0.61\pm 0.05$ and
&to determine the evolution\\
&(MG00) & &$\Log\ L^*=(28.13\pm 0.37)\, h^{-2}\,{\rm \ erg\ s}^{-1}{\rm \ Hz}^{-1}$
&\bigskip\\
\multirow{5}*{1.4GHz} &
&\multirow{5}*{Eq.~(\ref{eq:condonfunc})}
&\multirow{5}*{\vbox{\hbox{$ Y=2.85,\ X=22.40,\ W=2/3,\ B=1.5$}%
\hbox{\ (assuming $H_0=50$)}}}
&$\varphi(z)\propto (1+z)^{-0.0579}$ \\
&Condon\,(C89)\kern-1em& & &$\qquad\cdot\exp\left[ -\left( \frac{z}{14.3} \right)^{23.1} \right]$ \\
&$+$ Haarsma & & &{\hspace{3em}\em with}\\
&et al.\ (H00) & & &$L(z)\propto (1+z)^{3.97}$ \\
& & & &$\qquad\cdot\exp\left[ -\left( \frac{z}{1.39} \right)^{1.02} \right]$ 
\bigskip\\
\multirow{3}*{1.4GHz} &\multirow{3}*{\vbox{\hbox{Sadler}\hbox{et al.\ (S02)}}}
&\multirow{3}*{Eq.~(\ref{eq:saundersfunct})}
&$\varphi^*=(7.8\pm0.7)        \e{-2}\, h^3\ {\rm Mpc}^{-3} {\rm dex}^{-1},$
&constraints are too loose\\
& & &$ \alpha=0.84 \pm 0.02, \sigma=0.940\pm 0.004$ and
&to determine the evolution\\
& & &$\Log\ L^*=(27.15\pm 0.03)\, h^{-2}\,$ \ergsHz
&\bigskip\\
Blue &Wolf et al.\ (W03) &\multicolumn{3}{l}{15 Schechter functions with different parameters
according to spectral type and redshift bin}
\smallskip\\
\hline
  \end{tabular}
\end{table*}

\subsection{The mid infrared LF}
\label{sec:MIR}

We consider the recent determination of the mid-infrared (MIR) LF by
\citet[][hereafter P04]{pozzi04} with ISOCAM data at $15\mu$ from the
ELAIS survey. The local LF was derived from a sample of 150 objects
with $z\le 0.4$ and flux $S_{15\mu}\gtrsim 1$ mJy observed in two
fields for a total area of $2.1$ square degrees. P04 divided the
galaxies between quiescent spirals and starburst on the basis of the
MIR to optical flux ratio. The best-fit LF is given as two Saunders
functions (Eq.~\ref{eq:saundersfunct}) with the parameters shown in
Table~\ref{tab:LFs}.
We converted these LFs to the $60\mu$ wavelength using the
observed relation $L_{60\mu}/L_{15\mu}\sim 5$ from \citet{mazzei01}. 
The sum of the two $60\mu$ LFs is shown in
Fig.~(\ref{fi:infratake}) as the dot-dashed line.  The grey area shows
the $1\sigma$ error bar derived from a $1/V_{\rm max}$ analysis.
This local LF matches well the IRAS ones in the region around the knee
(i.e.\ $10^{10}\lesssim L\lesssim 10^{11}\ L\sun$), while at lower and higher
luminosities it stays a factor of two--three lower (see discussion in
P04); the evolution parameters were determined both by
analysing the LF in the redshift interval in which it is defined and
by fitting the MIR source counts.  P04 found that the
evolution of the MIR LF can be described as follows: no evolution
for the spirals from $z=0$ to $\sim 0.4$, and {\em both} luminosity
and density evolution for the starbursts with $z\leq 1$ (no evolution
for the starbursts beyond redshift 1). The best fit parameters for
evolution are $\eta_{\rm l}=3.5^{+1.0}_{-3.5}$ and $\eta_{\rm
d}=3.8\pm 2$.

\subsection{The radio LFs}
\label{sec:radiolf}

\begin{figure}[tp]    
  \centering
  \includegraphics[width=\columnwidth]{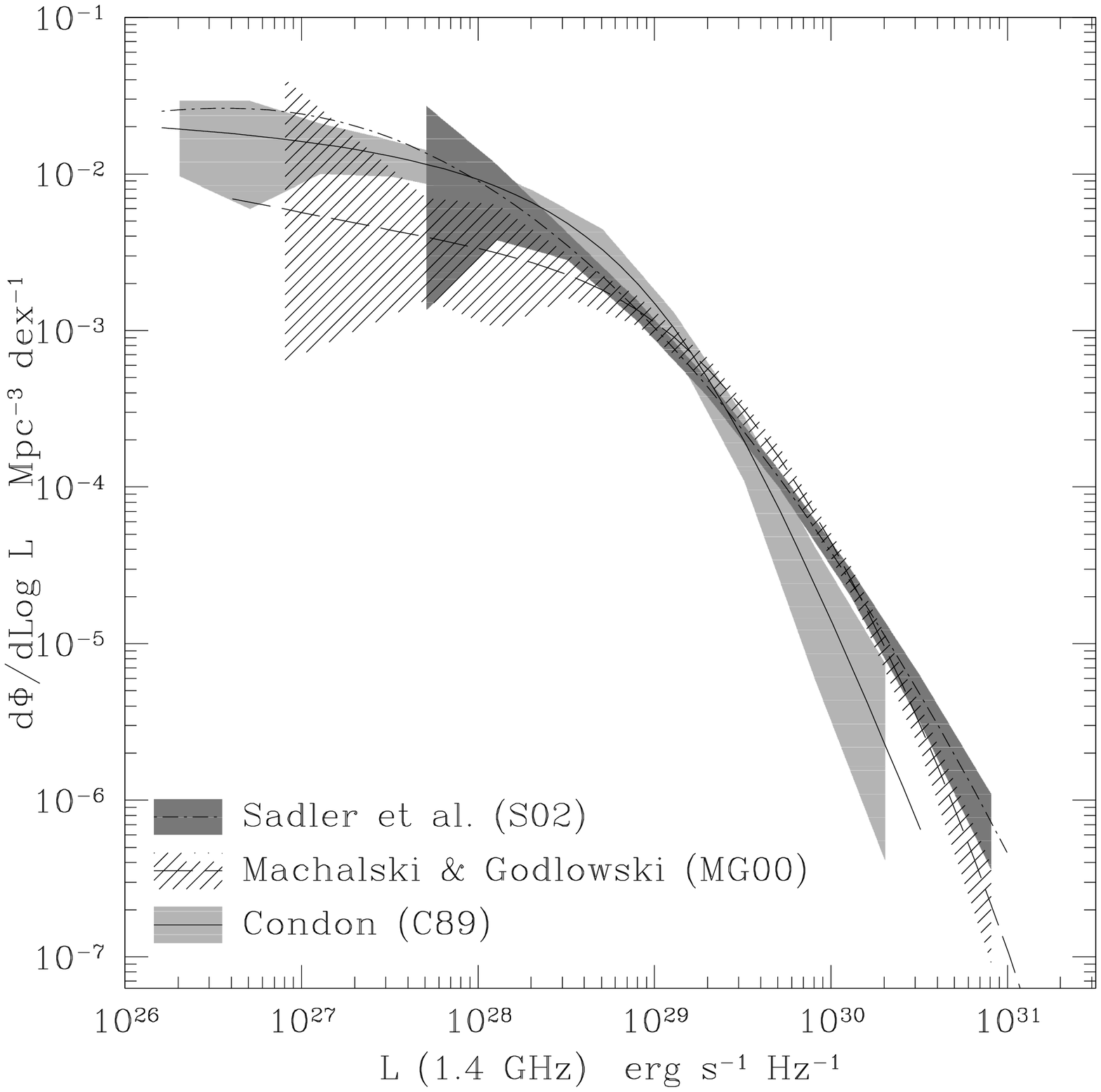} 
  \caption
  { Radio local luminosity functions. Solid curve: LF from spiral
  galaxies in the RSA (C89);  light grey area: error on the
  RSA spirals LF from $1/V_{\rm max}$ analysis.  Dot-dashed line:
  radio LF from star forming galaxies in both the 2dF Galaxy Redshift
  Survey and the RSA spirals (S02);  dark grey area: errors
  from $1/V_{\rm max}$ for the 2dFGRS galaxies.  Dashed curve and
  diagonally shaded area: LF for starburst galaxies in the Las
  Campanas Redshift Survey (MG00) and corresponding
  $1/V_{\rm max}$ error.  $H_0=70$ is assumed.  \label{fi:radiolf} }
\end{figure}

Among the first determinations of the radio (1.4 GHz) LF of galaxies,
the LF by \citet[][hereafter C89]{condon89} was derived by considering
the radio observations of 292 spiral galaxies with blue magnitude
$B\le12$ and declination north of $\delta=-45\degr$ from the Revised
Shapley-Ames (RSA) catalogue \citep{RSA}.  The C89 LF, shown in
Fig.~(\ref{fi:radiolf}) as the solid line, is parameterised as
\begin{eqnarray}
  \label{eq:condonfunc}
\Log\,\varphi(\Log L) &= &28.83 +Y-1.5\,\Log L  \nonumber \\
&&\kern-4em  - \left( B^2+\frac{(\Log L-X)^2}{W^2}
  \right)^{1/2} \quad {\rm Mpc}^{-3}\ {\rm dex}^{-1} .
\end{eqnarray}
with the best-fit parameters shown in Table~\ref{tab:LFs}. In addition
C89 considered also a sample of 142 galaxies selected
from the Uppsala General Catalogue of Galaxies \citep[UGC,][]{ugc}
with declination $-2.5\degr\le\delta\le 82\degr$, angular diameter
$\ge 1.0^\prime$ and classified as ``starburst'' or AGN on the basis
of their radio morphology, infrared/radio and $60\mu/25\mu$ flux
ratios. It was found that the LF parameters derived for the RSA
spirals fitted well also the UGC starbursts.

The most recent determinations of the radio LF of star forming
galaxies are based on the cross correlation of galaxy redshift surveys
at optical wavelengths with radio all-sky surveys. First we consider
the sample of radio detected galaxies in the Las Campanas Redshift
Survey \citep{machalski99}, which comprises 1157 galaxies with
magnitude $R\leq 18.0$ detected in the NRAO VLA Sky Survey (NVSS),
i.e.\ brighter than 2.5 mJy beam$^{-1}$ at 1.4 GHz: 502 galaxies were
classified as ``starburst'' and 655 as ``AGN'' on the basis of their
FIR-radio flux ratios, 25$\mu$--60$\mu$ spectral indexes and
radio-optical flux ratios. \citet[][hereafter MG00]{machalski00}
derived the LFs for both classes of objects; here we consider only the
``starbursts'' (Table~\ref{tab:LFs}; dashed line in
Fig.~\ref{fi:radiolf}).

The redshifts of the Las Campanas ``starburst'' galaxies are in the
range $0.01\leq z\leq 0.075$. As in the case of the IRAS galaxies,
this redshift range is too small to allow a reliable estimate of the
evolution. The constraints on the evolution found by MG00 are too
loose to either confirm or reject cosmic evolution for the Las
Campanas galaxies.

We also consider the radio LF of galaxies from the 2dF Galaxy Redshift
Survey (2dFGRS; \citealt{sadler02}, hereafter S02): 242 galaxies with redshift
$z<0.3$ were identified in the cross-correlation of 2dFGRS sources
(covering an area of 325 deg$^{-2}$) with the NVSS. The galaxies were
classified as star forming systems (different from AGN) on the basis
of their optical spectra. To further extend the LF at luminosities lower
than $\sim 10^{29}$ \ergsHz, S02 included also the RSA
spirals previously considered by C89.  The LF was
parameterised as a Saunders function, with the best fit parameters
shown in Table~\ref{tab:LFs}; it is the dot-dashed line in
Fig.~(\ref{fi:radiolf}). A V/V$_{\rm max}$ test for evolution of the
2dFGRS galaxies yielded inconclusive results, mainly because of poor
statistics (at the flux limit of the NVSS, only the highest luminosity
bins of the LF are sampled for $0.2\lesssim z\lesssim 0.3$).

The behaviour of the radio LF at higher redshifts ($z\lesssim 2$) was
explored by \citet{haarsma00}, who built a sample of galaxies detected
in deep radio surveys. They considered the C89 local LF and fitted its
evolution to the high redshift data.  \citet{haarsma00} considered a
rather complex form for the evolution with six free parameters (see
Table~\ref{tab:LFs}). However, their best-fit evolution parameters
closely resemble a pure luminosity evolution not too far from the one
found by T03, with the added constraint of an exponential
cutoff around $z\sim 1.4$. In the following, we shall refer with `C89
LF' to the C89 LF with the evolution determined in \citet{haarsma00}.

\subsection{The blue LF}
\label{sec:bluelf}

The evolutions of the infrared and radio LFs discussed in the previous
section are quite loosely constrained, either because of the limited
redshift range, or because of the limited number of objects present in
the surveys. However, larger surveys are currently ongoing at optical
wavelengths exploring wider redshift ranges. Here we consider the
COMBO-17 survey \citep[][hereafter W03]{wolf03} which provides a blue
(Johnson B-band) LF from 25,000 galaxies with $0.2<z<1.2$. Redshifts
are obtained from photometric observations taken with a set of 17
filters. The foremost data analysis goal of the COMBO-17 approach is
to convert the photometric observations into a low resolution spectrum
that allows simultaneously a reliable spectral classification of
stars, galaxies of different types and quasars, as well as an accurate
redshift estimate.

The galaxy B-band LFs are given in W03 for five redshift
bins (from $z=0.2$ to 1.2), so that no assumptions have to be made
about evolution. The LFs are also given for four different spectral
types defined on the basis of the grid of template spectra by
\citet{kinney96}.  Since spectral types may be broadly matched to
morphological types \citep{madgwick02}, we consider the LFs (one for
each redshift bin) for late-type galaxies as derived by
W03, i.e.\ with the exclusion of spectral type 1 in their
classification (see Table~\ref{tab:LFs}).


\section{The predicted XLFs at $z=0$ and their evolution}
\label{sec:xlf-discussion}

The LFs discussed in \S~\ref{sect_FIR2XLF} are converted by
using the approach first developed in \citet{avnitan86} (see also
\citealt{ioannis99} and \citealt{colin}):
\begin{eqnarray}
\varphi_{\rm X}(\Log\ \LX) &= &\int\limits_{-\infty}^{+\infty}
\varphi_{\rm Y}(\Log\ L_{\rm Y})  \nonumber \\
& &\qquad \cdot P\left(\Log\ \LX|\Log\ L_{\rm Y}\right)
\de\Log\ L_{\rm Y} 
\end{eqnarray}
where $P\left(\Log\ \LX|\Log\ L_{\rm Y}\right)$ is the probability
distribution for observing $\LX$ for a given luminosity in the Y band.
In RCS03 it was reported that the X-ray luminosity is tightly
correlated with radio and FIR luminosities ($\Log L_{0.5-2} = \Log
L_{1.4\rm GHz} +11.08$ and $\Log L_{0.5-2} = \Log L_{60\mu} +9.05 $,
with $L_{0.5-2}$ in \ergs, and with $L_{1.4}$ and $L_{60\mu}$ in
\ergsHz).  By assuming a Gaussian probability distribution for these
correlations, one has
%
\begin{eqnarray}
P(\Log\ \Lsx|\ \Log\ L_{60\mu}) &= &\nonumber \\
& &\kern-4em \frac{1}{\sqrt{2\pi}\sigma} \,
 {\rm e}^{-\frac{{\rm Log}\ L_{60\mu} +9.05 - {\rm Log}\ \Lsx}{2\sigma^2}}
\end{eqnarray}
with $\sigma\sim 0.30$, and for the radio band
%
\begin{eqnarray}
 P\left(\Log\ \Lsx|\Log\ L_{1.4\rm GHz}\right) &= &\nonumber \\
& &\kern-5em \frac{1}{\sqrt{2\pi}\sigma} \,
 {\rm e}^{-\frac{{\rm Log}\ L_{1.4} +11.10 - {\rm Log}\ \Lsx}{2\sigma^2}}
\end{eqnarray}
with $\sigma\sim 0.24$.

To convert the W03 LF into an XLF we consider the B-band/X-ray
correlation, which has been the subject of much attention in X-ray
studies of galaxies of all morphological types
\citep{giacconi87,fabbiano89,djf92,shapley01,fabshap02}.  A
correlation with the form $\LX\propto L_{\rm B}^{1.1^{+0.1}_{-0.2}}$
was first found by \citet{fgt88} by analysing 51 spiral galaxies
observed with {\em Einstein}.  An enlarged sample of 234 spirals from
the {\em Einstein} archive was built by \citet{shapley01}, from which
we derive $\LX\propto L_{\rm B}^{1.13\pm 0.09}$ by performing a
linear regression analysis \citep{isobe90} with the Schmitt method in
the ASURV code for censored data analysis \citep{asurv}, which we
chose because of the large number of upper limits on the X-ray
luminosities and consistently with \citet{shapley01}.
Thus we use the following relationship
\begin{eqnarray}
P\left(\Log\ \Lsx|\Log\ L_{\rm B}\right) &= &\nonumber\\
& &\kern-4em \frac{1}{\sqrt{2\pi}\sigma} \,
{\rm e}^{-\frac{1.13{\rm Log}\ L_{\rm B} -8.97 - {\rm Log}\ \Lsx}{2\sigma^2}}
\end{eqnarray}
with $\sigma\sim 0.54$.

We note that while the B/X-ray correlation was derived by making use
of magnitudes corrected for extinction, caused by the inclination of
the galaxy on the plane of the sky, the W03 LF is not corrected for
extinction. However, this should not be a serious problem, since the
median inclination-induced extinction for the galaxies in the Third
Reference Catalogue of Galaxies \citep[RC3,][]{rc3} is 0.26 magnitudes
(i.e.\ a factor of $\sim 1.3$), so that we have to correct the LF
for inclination-induced extinction by rescaling the normalisation of
the LF by a factor 1.3.


%

\begin{figure}[tp]    
   \includegraphics[width=\columnwidth]{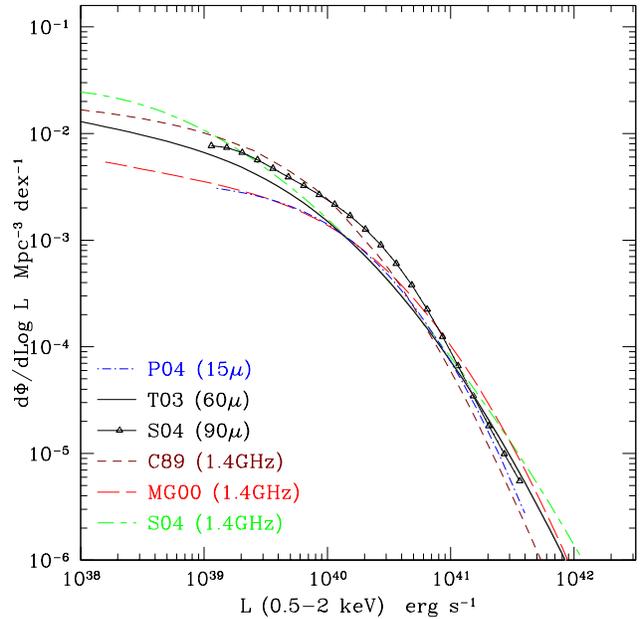}
    \caption { \figuraacolori
      IRAS, ISO, and radio local luminosity
      functions converted to
      the X-rays. Solid (black) curve: IRAS 60$\mu$ LF from T03.
      Continuous line with triangles: ISO 90$\mu$ LF (S04).
      Dot-dashed (blue) curve: ISO 15$\mu$ LF (P04).
      Long-dashed (red) curve: radio LF (MG00). 
      Short-dashed (dark red) curve: radio LF (C89).
      Short-dash-long-dash (green) curve: radio LF (S02).
      \label{fi:xraytake} }
\end{figure}

The radio and IR derived XLFs are shown in
Fig.~(\ref{fi:xraytake}). Around the `knee' (i.e.\ $10^{40}\lesssim
\LX\lesssim 10^{41}$) all the predicted XLFs agree within a factor of
2, which is the approximate size of the error bars on the radio and
infrared LFs. Much more scatter is present among the tails of the LFs,
due to the small number statistics.

The main point here is that the infrared and radio LFs define a {\em
  locus} where an observed local XLF should fall. This confirms the
effectiveness of the various selection criteria of galaxies used in
the surveys. The X-ray luminosity density (XLD) of galaxies in the
local universe (${\rm XLD}= \int_{-\infty}^{+\infty} \de\Log L \
\varphi(\Log L) \cdot L$) is also well defined, with estimated values
that are spread within a factor $\sim 1.7$ around a formal average
$\sim 3.4\e{37}$ \ergs\ Mpc$^{-3}$; the T03 XLD being close to the
average value (Table~\ref{tab:luminositydensity}). Thus the T03 XLF,
built on the largest sample of galaxies, appears well suited to be
representative of the local XLF (moreover, had we weighted the mean by
using the number of galaxies in surveys, the average XLD would be
coincident with the T03 one).

%
%
%

We also computed the percentage contribution to the cosmic X-ray
background (CXB) in the 0.5--2.0 keV energy band by adopting the mean
XLD and a uniform distribution (no evolution) of the galaxies up to
$z=4$; as regards the CXB we have adopted the spectral shape of $9.8
E^{-1.4}$ photon s$^{-1}$ cm$^{-2}$ sr$^{-1}$ keV$^{-1}$, with
normalisation taken from \citet{revnivtsev05}. By adopting the
evolution parameters as discussed in \S~\ref{sect_obslogn} we find
that the contribution to the CXB is comprised in the percentage
interval $6\%$--$12\%$, keeping in mind, however, that the average
spectral slope of the galaxies is much softer than that of the adopted
extrapolation of the CXB. These findings are consistent with the $1\%$
contribution to the XRB by galaxies detected in the CDFs
\citep{hornsch03}.

\begin{table}[tp]
  \centering
  \caption{X-ray luminosity densities of galaxies at $z=0$ in the
    0.5--2.0 keV band.}
  \label{tab:luminositydensity}
  \begin{tabular}{lc}
\hline\hline
Luminosity function  &Predicted luminosity density\\
                     &(\ergs\, Mpc$^{-3}$)     \\
\hline
\citeauthor{takeu03} (T03)      &$\quad 3.0\e{37}$\\
\citeauthor{condon89} (C89)     &$\quad 4.1\e{37}$\\
\citeauthor{sadler02} (S02)     &$\quad 3.7\e{37}$\\
\citeauthor{pozzi04}  (P04)     &$\quad 2.5\e{37}$\\
\citeauthor{serje04}  (S04)     &$\quad 4.1\e{37}$\\
\citeauthor{machalski00} (MG00)\kern-2em &$\quad 2.9\e{37}$\\
\hline
average value              &$\quad 3.4\e{37}$\\
\hline
  \end{tabular}
\end{table}

\subsection{The evolution of the XLFs}
\label{sec:xlfevol}

\ifthenelse{\value{refereemode}=1}{
  \begin{figure*}[tp]    
    \centering
    \includegraphics[width=\columnwidth]{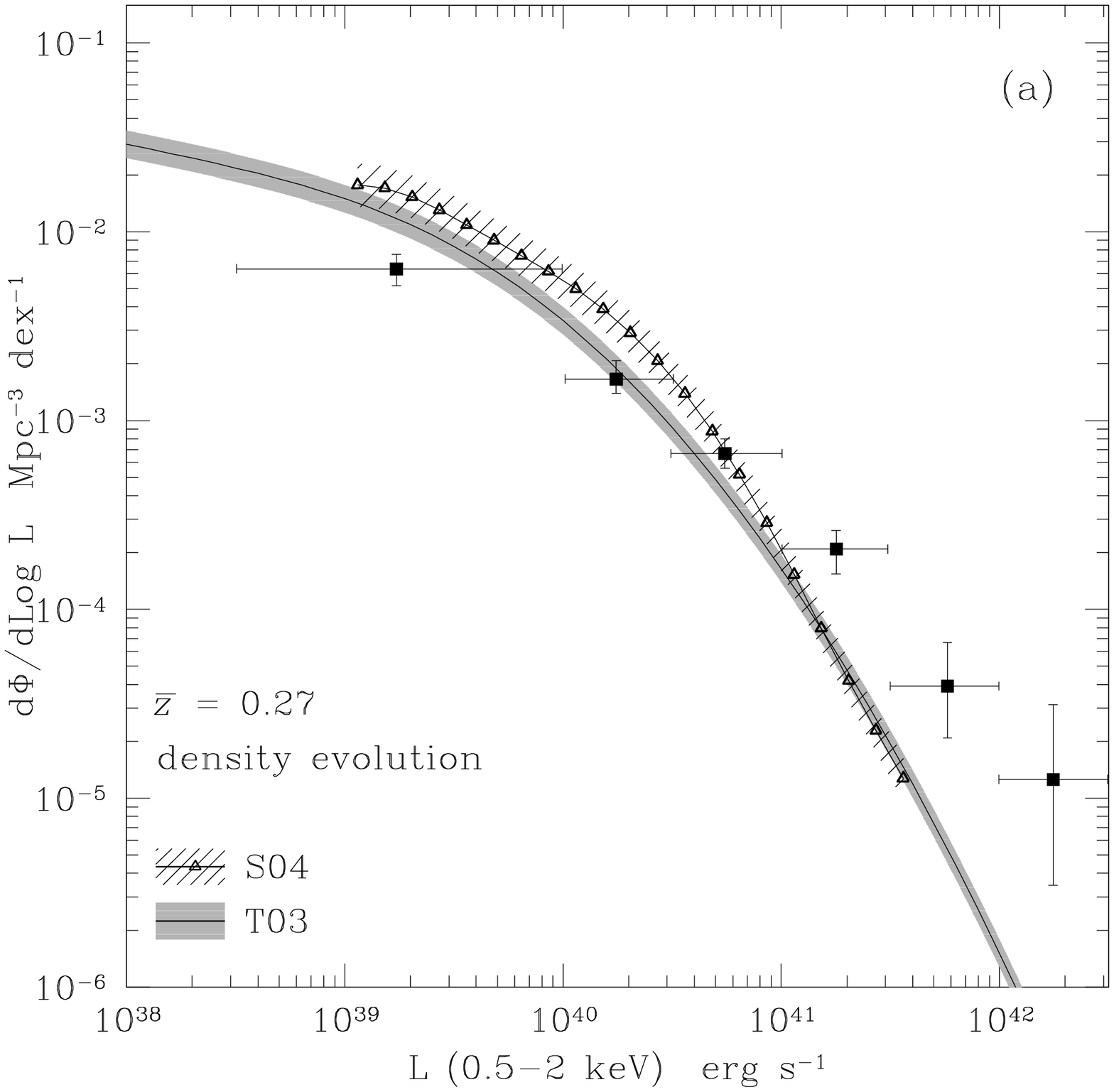}\\
    \includegraphics[width=\columnwidth]{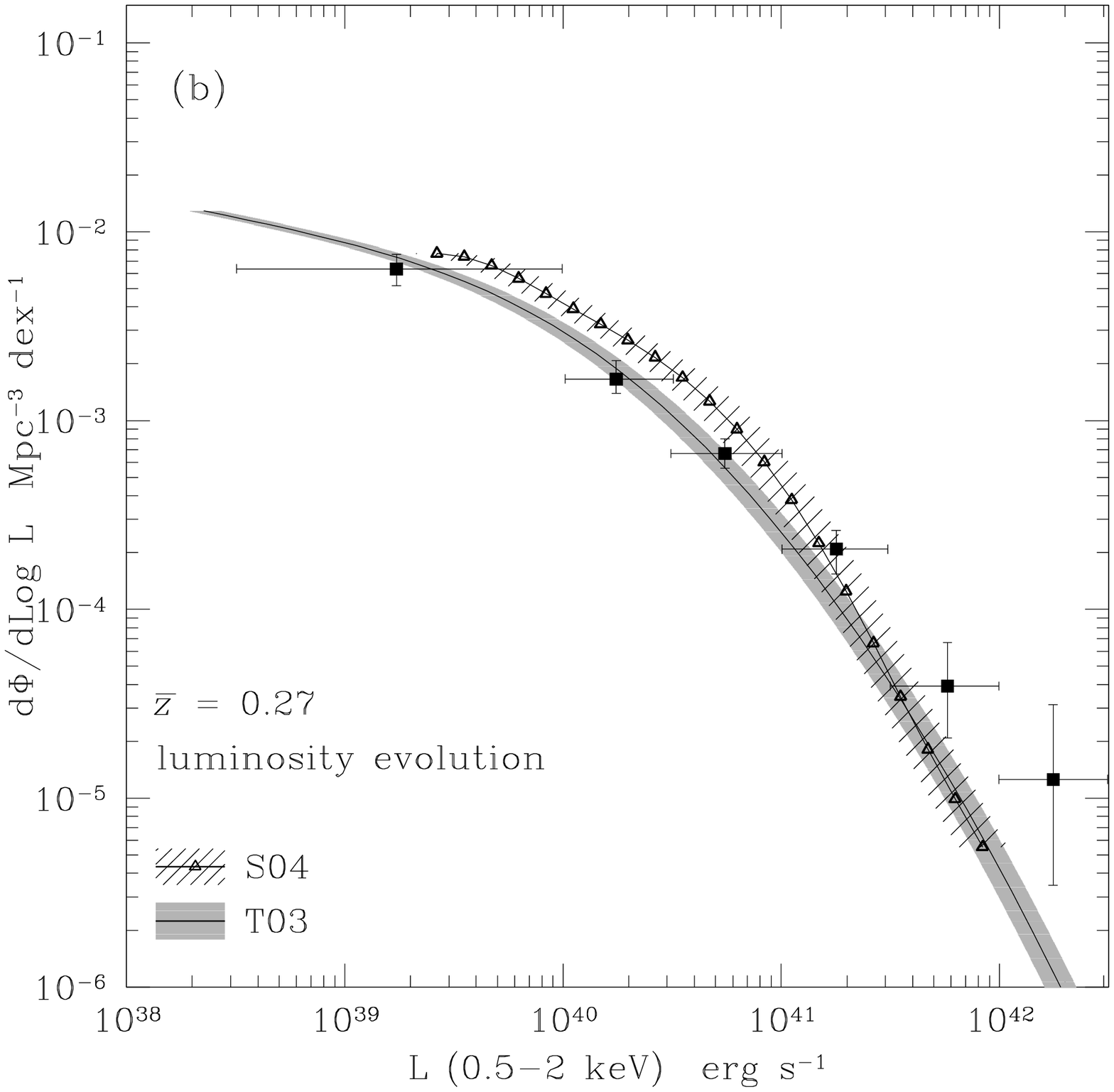}
  \end{figure*} 
  \begin{figure*}[tp]
    \centering
    \includegraphics[width=.8\columnwidth]{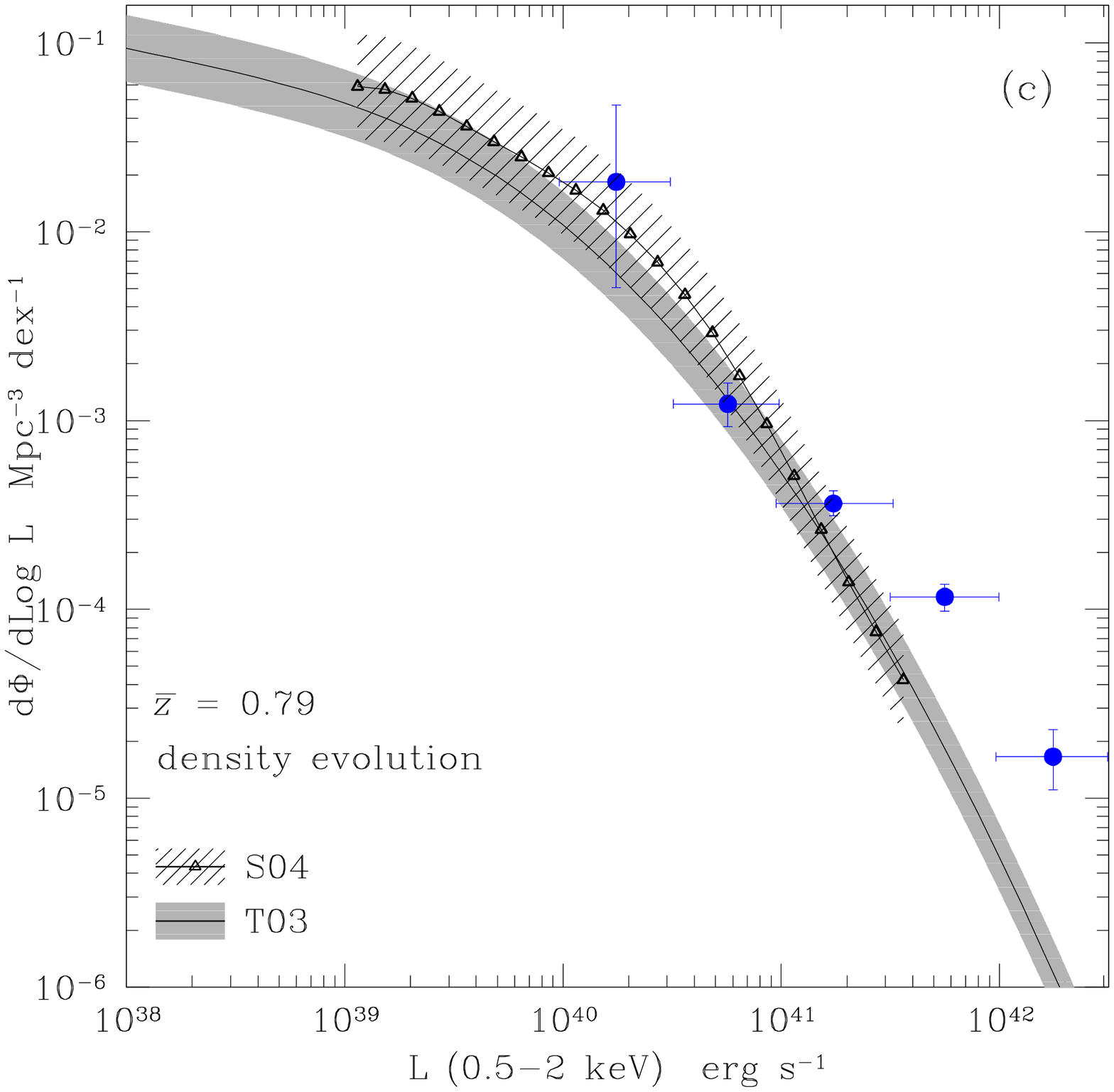}\\
    \includegraphics[width=.8\columnwidth]{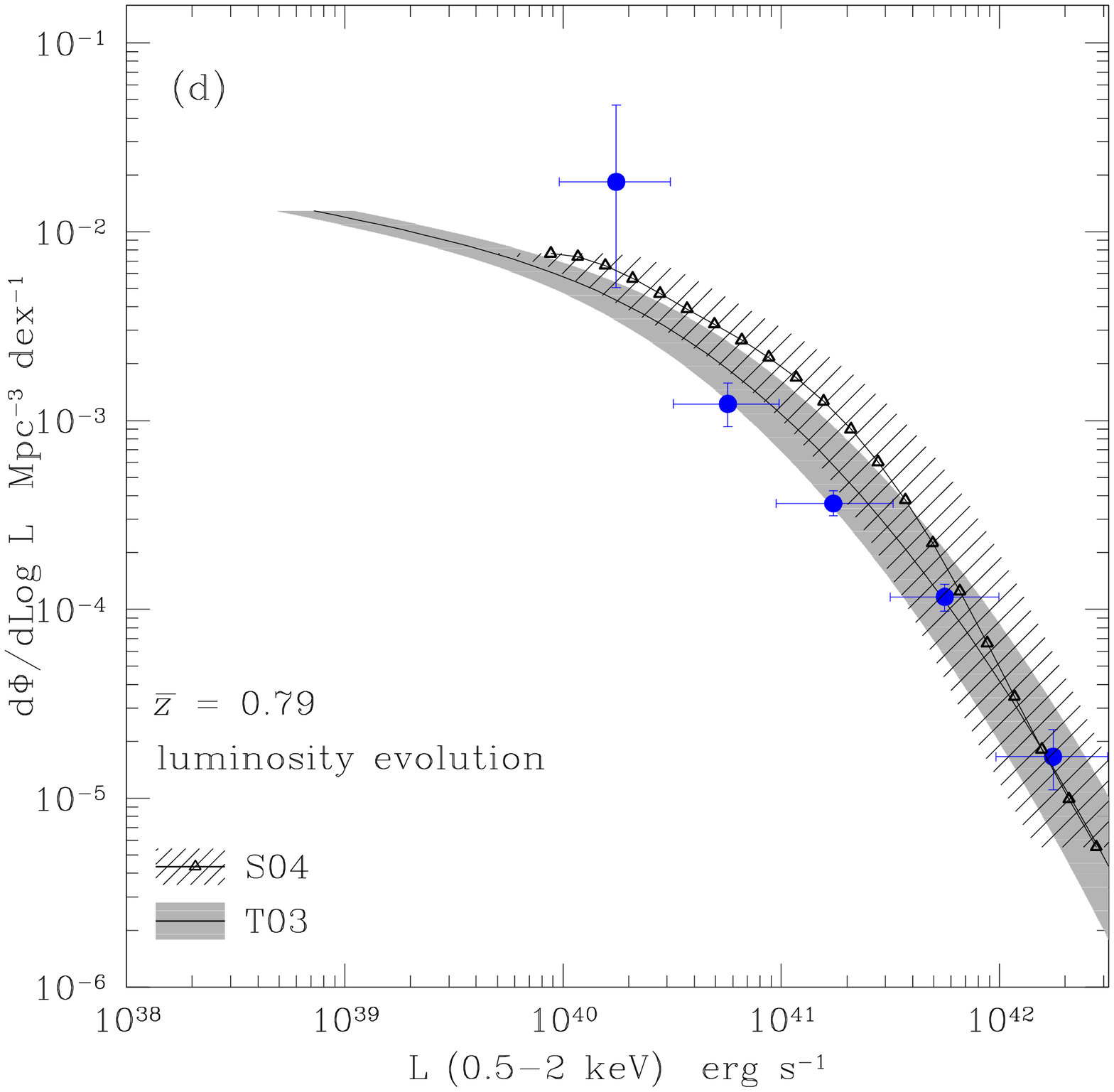}
}{
  \begin{figure*}[tp]
    \centering
    \includegraphics[width=0.497\textwidth]{lozdens.ps}
    \includegraphics[width=0.497\textwidth]{lozlum.ps}\\
    \includegraphics[width=0.497\textwidth]{hizdens.ps}
    \includegraphics[width=0.497\textwidth]{hizlum.ps}
}
  \caption { Comparison of the X-ray luminosity functions.  (a)
    Low-$z$ bin ($\bar z=0.27$), density evolution. (b) Low-$z$ bin
    ($\bar z=0.27$), luminosity evolution.  (c) High-$z$ bin ($\bar
    z=0.79$), density evolution.  (d) High-$z$ bin ($\bar z=0.79$),
    luminosity evolution.  Filled squares or circles: observed XLF
    from \citet{colin}. Solid curve: IRAS LF
    \citep{takeu03,takeu03err}; the grey shading shows the $1\sigma$
    error on the evolution exponent; triangles within line shading:
    ISO LF \citep{serje04}.  The XLFs have been evolved
    with the prescriptions listed in Table~\ref{tab:LFs} to the median
    redshifts of the two bins in which the observed XLFs were obtained.
    \label{fi:xlf-normanetal} }
\end{figure*}

\begin{figure*}[tp]    
\centering
\ifthenelse{\value{refereemode}=1}{
  \includegraphics[width=.8\columnwidth]{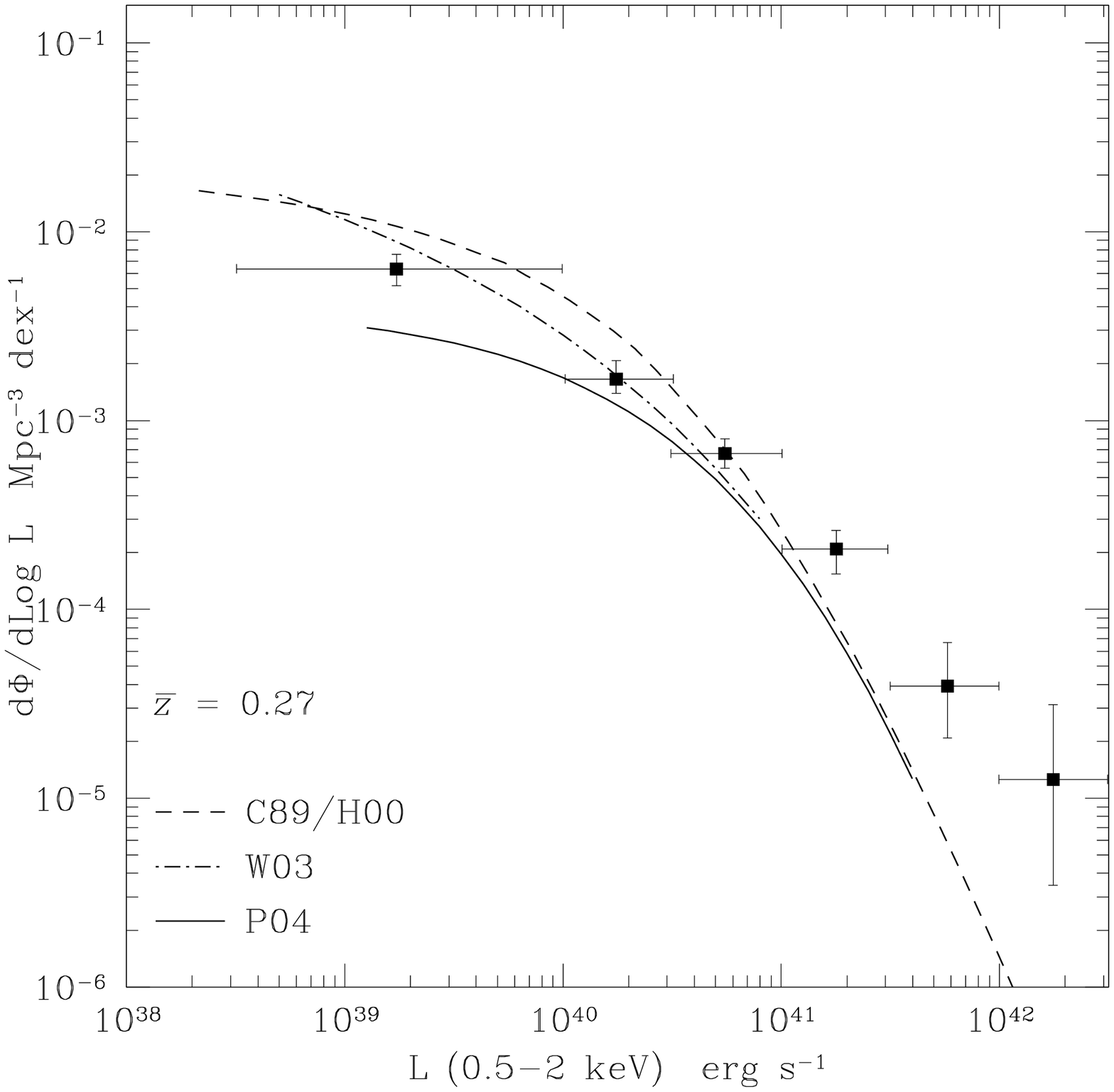}\\
  \includegraphics[width=.8\columnwidth]{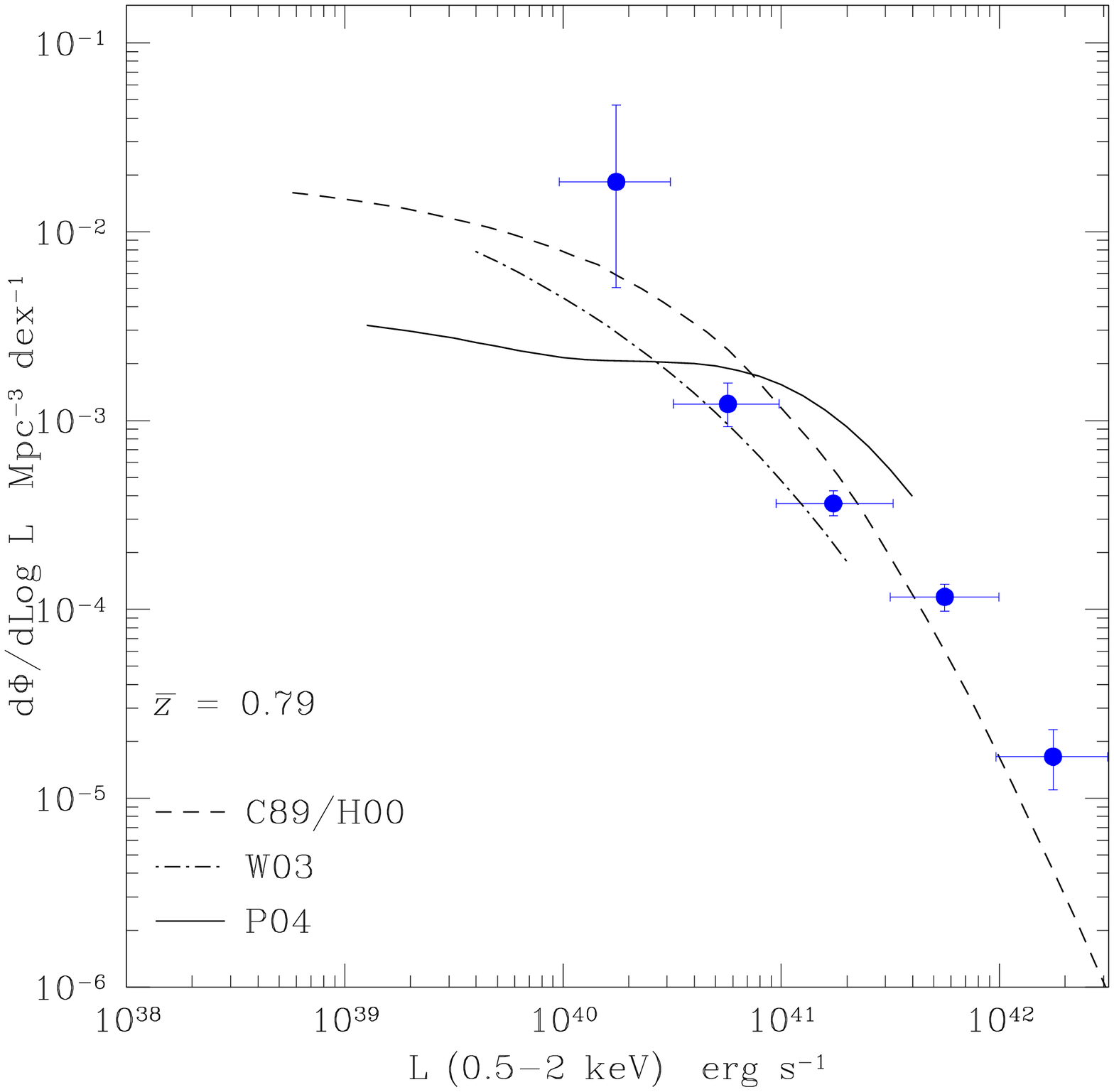}
}{
  \includegraphics[width=0.497\textwidth]{lozdue.ps}
  \includegraphics[width=0.497\textwidth]{hizdue.ps}
}
  \caption { Comparison of the X-ray luminosity functions.  {\em Left
    panel:} low-$z$ bin ($\bar z=0.27$).  {\em Right panel:} high-$z$
    bin ($\bar z=0.79$).  Data points: observed XLFs from
    \citet{colin}. Solid curve: ISO XLF (P04), with $1\sigma$ error on
    evolution (see text); dashed curve: radio XLF (C89); dot-dashed
    curve: blue XLF (W03), observed in the redshift bins
    0.2--0.4 and 0.6--0.8, respectively.  The MIR and radio XLFs have
    been evolved 
    with the prescriptions listed in Table~\ref{tab:LFs} to the
    median redshifts of the two bins in which the observed XLFs were
    calculated. 
    \label{fi:xlf-normanetal2} }
\end{figure*}

The infrared-, radio- and blue-derived LFs are compared with the X-ray
observed LF in Figs.~(\ref{fi:xlf-normanetal}) and
(\ref{fi:xlf-normanetal2}), where all the LF are drawn only in the
X-ray luminosity range corresponding to that spanned by the data
defining each LF in the original band.  Since the \citet{colin} XLF
was derived at higher redshifts ($\bar z\sim 0.27$ and $\bar z\sim
0.79$), in order to perform a meaningful comparison we have to
consider the different constraints and parameterisations on the
evolution of the LFs.  The LFs whose evolutions are constrained only
by local data ($z\lesssim 0.3$, i.e.\ the T03 and S04 LFs) are shown
in Fig.~(\ref{fi:xlf-normanetal}).  Only pure density evolution was
considered for these LFs by their respective authors.  Nonetheless, it
seems that pure luminosity evolution may provide an equally good fit
to the data, at least for the T03 LF for which the analysis is ongoing
(Takeuchi, priv.\ comm.).  However, we have decided to explore also
the effect of pure luminosity evolution by adopting the same exponents
reported for the density evolution (this is not formally correct but
may serve as a first-order approximation).  The other LFs (P03, H00
and W03), whose evolutions are constrained from higher redshift data,
are shown in Fig.~(\ref{fi:xlf-normanetal2}).  These are either
parameterised in both luminosity and density (P04 and H00), or are
{\em observed} LFs in given redshift bins (W03).  Two LFs are not
further discussed: \citet{saun90}, because it is indistinguishable
from the T03, and the S02 and MG00 LFs because the constraints on
their evolution are too loose to add significant information to our
discussion.

In both figures the shaded areas show, where available, the $1\sigma$
error on evolution, while the errors on the $z=0$ LF are ignored.


Considering pure density evolution (Figs.~\ref{fi:xlf-normanetal}a and
\ref{fi:xlf-normanetal}c), we find that the X-ray derived LF is in
agreement with the FIR- and radio-derived XLFs for $L_{\rm X}\lesssim
10^{41.5}$, but it is significantly flatter at its bright end ($L_{\rm
X}\gtrsim 10^{41.5}$), where it has a density of galaxies which is
about one order of magnitude larger than that of the FIR-radio
galaxies.  This discrepancy is partly removed when we consider pure
luminosity evolution (Figs.~\ref{fi:xlf-normanetal}b and
\ref{fi:xlf-normanetal}d), as it remains only noticeable (panel b) for
low-redshift, high-luminosity ($\LX>10^{42}$ \ergs) galaxies.
Although this discrepancy is only within $2\sigma$, it may be
symptomatic of a residual fraction of low luminosity AGN embedded in
the galaxies at the brightest end of the X-ray sample used by
\citet{colin}.

The XLF derived from the MIR LF (P04) appears to lie below the
observed XLF in the low redshift bin (Fig.~\ref{fi:xlf-normanetal2},
left panel) and to cross the XLF in the high redshift bin (same
figure, right panel). This might be partly explained because of the
decoupling between the evolution of quiescent and actively star
forming galaxies: the net effect is that only the high luminosity
($L_{\rm X}\gtrsim 10^{40}$ \ergs) tail of the LF evolves, but the
non-evolving, lower luminosity end of the LF cannot fit the lowest
luminosity, low redshift point of the observed XLF.

The XLF derived from the radio LF (H00) crosses within $1\sigma$ most
of the observed XLF data points, with the exception of the two highest
luminosity points in the low redshift bin
(Fig.~\ref{fi:xlf-normanetal2}, right panel).

The XLF derived from the blue LF (W03; dot-dashed curve in
Fig.~\ref{fi:xlf-normanetal2}) gives results which are very similar to
the T03 XLF.

\begin{figure}[tp]    
\centering
  \includegraphics[width=.8\columnwidth,bb=17 143 382 692,clip]{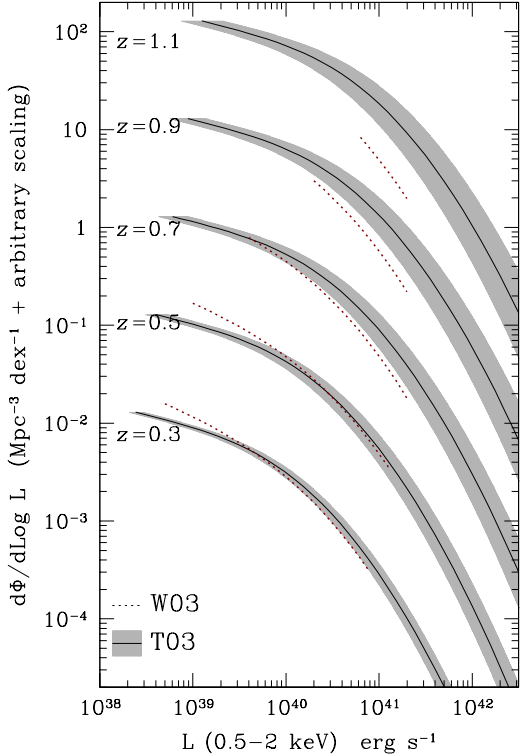} 
\caption { %
Comparison of the T03 and W03 XLFs for each redshift bin in which the
W03 LFs are defined. Pure luminosity evolution with $\etalum=3.4\pm
0.7$ is assumed for the T03 XLF.  Solid line with grey error area:
T03, with $1\sigma$ error on evolution. Dotted line: W03.  The XLFs
are progressively shifted for clarity by 1 dex along the vertical
axis.  The agreement is quite good in the low redshift bins and
worsens at higher $z$. A better agreement is found if one
assumes $\etalum=2.7$ for the T03 XLF, corresponding to the lower
bound of the grey error area.
\label{fi:xlf-takeucombo5} }
\end{figure}

Thus, it seems that pure luminosity evolution is somewhat favoured
over pure density, with $\etalum$ between 2.7 (the best fit value
found by \citealt{colin}) and 3.4 (adopted from T03).

The W03 XLF offers the possibility to check the evolution
independently from the \citet{colin} XLF and in a larger number of
redshift bins. Moreover, because of the large size of the
galaxy sample and the wider redshift range, the luminosity and density
evolution can be analysed separately. W03 reports on the evolutionary
patterns for each spectral type: on the whole, these patterns can be
understood as an increase in starburst activity with redshift, with an
activity maximum at $z\gtrsim 1$, in combination with a propagation of
the galaxies through the spectral types as the mean age of the stellar
population decreases. 
Quantitatively, we may consider the redshift evolution of both the
X-ray luminosity density and of the number density of the W03
galaxies. We find that the luminosity density doubles from $z\sim 0.3$
to 1.1, while the number density stays almost constant. This pattern
is consistent with luminosity evolution prevailing over density.

  A detailed comparison of the W03 XLFs with the evolved
T03 XLF (pure luminosity, $\etalum=3.4$) shows a fairly good agreement
(less than a factor of 2) in the redshift bins from $z\sim 0.3$ to
$\sim 0.7$, while it is a factor of $\sim 3$ below the T03 XLF in the
$z\sim 1.1$ bin (Fig.~\ref{fi:xlf-takeucombo5}).  No agreement would
occur had we considered pure density evolution. The assumption of pure
luminosity evolution for the T03 XLF, which was arbitrarily introduced
in \S~\ref{sec:xlf-discussion}, appears to be validated by this
comparison. We notice, however, that the evolved T03 systematically
exceeds the W03 XLF for $z\gtrsim 0.7$, perhaps an indication that the
$\etalum=3.4$ evolution is too strong; in fact, a milder exponent
(e.g.\ $\etalum=3$) would reconcile the T03 and W03 XLFs for
$z\lesssim 0.9$. In the $z\sim 1.1$ redshift bin the T03 XLF would
still exceed the W03 one by a factor of 2--3, possibly hinting that
the evolution should be stopped at $z\sim 1$.  Altogether the two by
far largest samples of sources provide a complementary and self
consistent picture up to a redshift $z\sim 1$, and in the analysis
performed in the following Sects.\ we shall only consider the T03 XLF.

\section{The X-ray \lognlogs of galaxies}
\label{sect_obslogn}

We now turn our attention to the analysis of the X-ray number counts
of galaxies, since they represent an immediate and alternate test bed
for both the selection criteria in X-ray surveys and for the shape and
evolution of different LFs.


\begin{figure*}[tp]     
  \sidecaption
  \includegraphics[width=12cm]{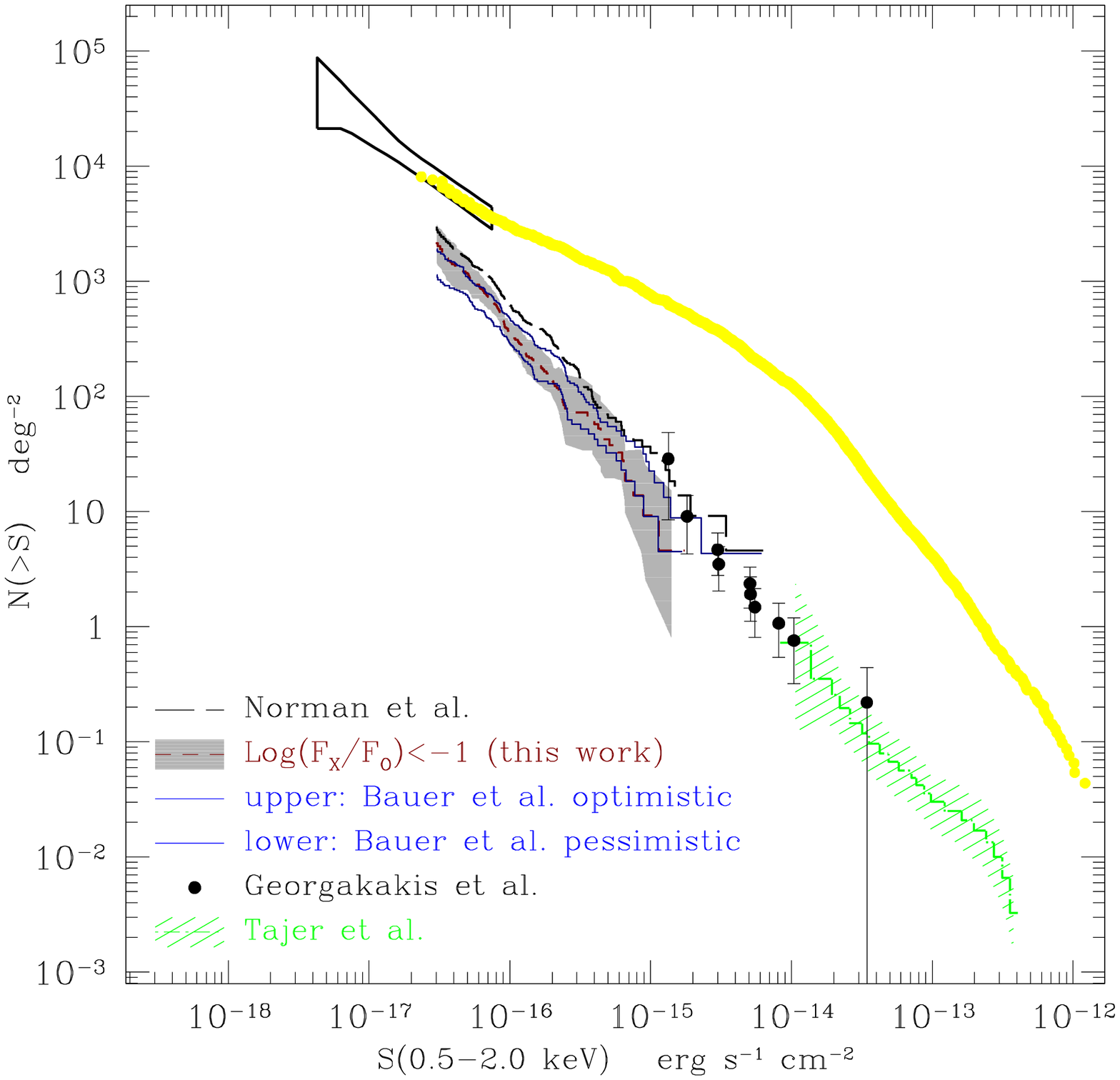}
  \caption { \figuraacolori Observed X-ray counts. The thick,
    upper line and the horn-shaped symbol show the observed \lognlogs
    for all X-ray sources in the \chandra\ Deep Fields
    \citep{moretti03} and the limits from the fluctuation analysis
    \citep{miyaji02a,miyaji02b}. 
    Short-dashed red histogram and grey area: sources with Log
    (X-ray/optical flux ratio) $<-1$ with Poissonian error (this
    work). The width
    of the error area may be taken as representative for the following
    \lognlogs too.  Long-dashed black (upper) histogram: sources from
    the Bayesian sample of \citet{colin}.  Continuous blue histograms:
    `optimistic' (upper) and `pessimistic' (lower) estimates from
    \citet{bauer04}. The data points and the dot-dashed green
    histogram at fluxes $\gtrsim 10^{-15}$ \ergscmq\ show the
    \lognlogs of serendipitous spiral galaxies in \xmm\
    \citep{georgakakis04} and ROSAT HRI fields \citep{tajer05}, respectively.
    \label{fi:xcts-baueretal} }
\end{figure*}

\subsection{The observed \lognlogs}

The normal galaxy number counts at faint fluxes have been presented
several times in the literature. While we just mention the first
prediction based on optical counts by \citet{giacconi87}, we focus
here on the different determinations of the galaxy \lognlogs based on
\chandra\ Deep Fields catalogues, which are defined at fluxes
$3\e{-17}\lesssim S_{0.5-2\rm keV}\lesssim 10^{-15}$ \ergscmq.
  
The first determinations were made by \citet{bauer02}, a sample of 11
galaxies selected by cross correlating X-ray and radio sources with
optical spectra classified as ``emission line galaxies'' in the inner
$3^\prime$ of the CDFN, and by \citet{hornsch03}, a sample of 43
galaxies with $\fxott<-2.3$ (a more stringent criterion to avoid AGN
contamination). \citet{hornsch03} found also that the fraction of
early type systems in their sample is somewhat low ($\sim 15\%$); thus
we believe that in the following analysis the contribution form early
type galaxies may be safely ignored.

By refining the selection criteria it has been possible to construct
the \lognlogs for larger samples to which we shall refer in our
discussion. These are:

\begin{itemize}
\item The \lognlogs of the Bayesian sample defined in \citet{colin}
  and discussed in \S~\ref{sect_norman}. By considering the sky
  coverage of the \chandra\ Deep Field North (CDFN) survey, as
  reported in \citet{alexander03}, we derived the \lognlogs for these
  sources, which is shown in Fig.~(\ref{fi:xcts-baueretal}) as the
  long-dashed histogram;
\item Two samples defined in \citet{bauer04}. The first one comprises
  175 sources with internal absorption $N_{\rm H}<10^{22}$ cm$^{-2}$
  and hardness ratio HR$<-0.8$ and X-ray luminosity $L_{0.5-8 \rm
    keV}<3\e{42}$ \ergs, or with off nuclear emission\footnote{A
    source is catalogued as `off nuclear' if point-like X-ray emission
    is found within the optical contour of a galaxy in an off-centre
    position.}. The second one comprises 109 sources with $\fxott<-1$
  and $N_{\rm H}<10^{22}$ cm$^{-2}$ and HR$<-0.8$ and $L_{0.5-8 \rm
    keV}<3\e{42}$ \ergs\ and no broad lines in the optical spectrum.
  Because of their different sizes, the two samples are named as
  `optimistic' and `pessimistic' estimates, respectively
  (Fig.~\ref{fi:xcts-baueretal}).  These samples also represent an
  update of the \citet{hornsch03} analysis (Hornschemeier, priv.\
  comm.).
\item A sample of 142 sources from the CDFN and the \chandra\ Deep
  Field South, which we selected for having an X-ray/optical flux
  ratio $\lesssim -1$.  Although this selection criterion may be
  regarded as somewhat crude, it is indeed useful since it allows an
  immediate object selection with the least possible number of
  parameters.  The resulting \lognlogs is plotted as the short-dashed
  histogram in Fig.~(\ref{fi:xcts-baueretal}). An estimate of the
  Poissonian error (after \citealt{gehrels86}, with confidence level
  of $68.3\%$) for this \lognlogs is also shown as the grey area. The
  width of the error area may also be taken as representative for the
  other \lognlogsa.
\end{itemize}

We believe that our choice, although not exhaustive, characterises
both the scatter among different number counts determinations and the
`state of the art' selection criteria.  We find that there is a quite
good agreement between the slopes of the observed counts, while the
normalisations are comprised within a factor $\sim 2$.

These results are somewhat strengthened by the recently determined
galaxy counts at brighter X-ray fluxes ($S_{0.5-2}\gtrsim 10^{-15}$).
A smooth transition is observed between the bright tail of the number
counts of the galaxies in the deep fields and the recently determined
\lognlogs of galaxies with bright X-ray fluxes, as shown in
Fig.~\ref{fi:xcts-baueretal}: \citet{georgakakis04} selected 11
sources from the cross correlation of serendipitous X-ray sources in
archival \xmm\ fields with galaxies from the Sloan Digital Sky Survey
(SDSS); while \citet{tajer05} found 32 sources (from which we selected
21 galaxies with late-type morphology) by cross correlating a sample
of serendipitous X-ray sources in archival ROSAT HRI fields with
galaxies in the Lyon-Meudon Extragalactic DAtabase (LEDA).  The
observed integral counts appear to be consistent with one power law
(exponent $\sim -1.4$) spanning about four decades in flux.  For easy
reference in Figs.~(\ref{fi:xcts-takeu}--~\ref{fi:xcts-takeuptak}) we
always reproduce the counts obtained in the Chandra Deep Fields, but
omit the counts at the brighter hand that would require unnecessary
larger figures' size.  

\subsection{X-ray counts from integration of the LFs}
\label{sec:counts-integration}

The observed X-ray counts may be checked against the number counts
obtained by integrating the luminosity functions previously derived.
In the following, we shall only discuss the number counts obtained
from the T03 XLF, since it has already been shown that its properties
are the best defined ones and consistent with all the XLFs derived
from the other samples.

The number counts may be obtained by:
\begin{equation} 
N(>S) = \int\limits_{z_{\rm min}}^{z_{\rm max}}
\de z\!\!\!\!
\int\limits_{L_{\rm min}(z,S)}^{L_{\rm max}} 
\!\!\!\!\de \Log L \ 
\varphi(\Log L,z) \, \frac{\de V}{\de z} 
\end{equation}
where $\de V/\de z$ is the comoving volume between $z$ and $z+\de z$.
The K-correction can be safely neglected because the mean slope of the
X-ray photon spectra of star forming galaxies is $\Gamma\sim 2.1$
(RCS03).  We perform the integration in the luminosity interval
$10^{39}\leq \LX\leq10^{43}$ \ergs\ at $z=0$; when considering
luminosity evolution, the integration limits are also `evolved', i.e.\
they are shifted by $(1+z)^\etalum$. We note that, had we not evolved
the integration limits, the main conclusions would be unchanged; the
only noticeable difference would be a larger number (by a factor $\sim
3$) of very faint sources (flux $\lesssim 10^{-18}$ \ergscmq). We
perform the integration in the $0\leq z\leq 2$ redshift range by
assuming two different possibilities for the evolution: the full range
up to $z\leq 2$ and a restricted range up to $z\leq 1$.


In Fig.~(\ref{fi:xcts-takeu}, left panel) we plot the counts obtained
from the T03 XLF with density evolution ($\etaden=3.4$). The
integrated counts lie on the lower side of the \lognlogs bundle and
are consistent with the `pessimistic' estimate in \citet{bauer04}.

On the other hand, if we assume pure luminosity evolution with
$\etalum=2.7$ as determined in \S~\ref{sec:xlfevol}
(Fig.~\ref{fi:xcts-takeu}, right panel), the integrated counts would
be consistent with the observed ones, staying in the middle of the
observed counts bundle for $S_{0.5-2}\gtrsim 5\e{-17}$ \ergscmq.
Performing the luminosity evolution up to $z\sim 1$ or 2 would result
in a difference of less than a factor of 2 at fainter fluxes, with the
predicted counts at $3\e{-17}$ \ergscmq\ falling in the middle or in
the upper end of the bundle, respectively.  Had we assumed a stronger
exponent ($\etalum=3.4$ as explored in \S~\ref{sec:xlfevol}), the
predicted counts would have outnumbered the observed ones.  

\begin{figure*}[tp]    
  \begin{center}
  \ifthenelse{\value{refereemode}=1}{
    \includegraphics[width=0.72\columnwidth]{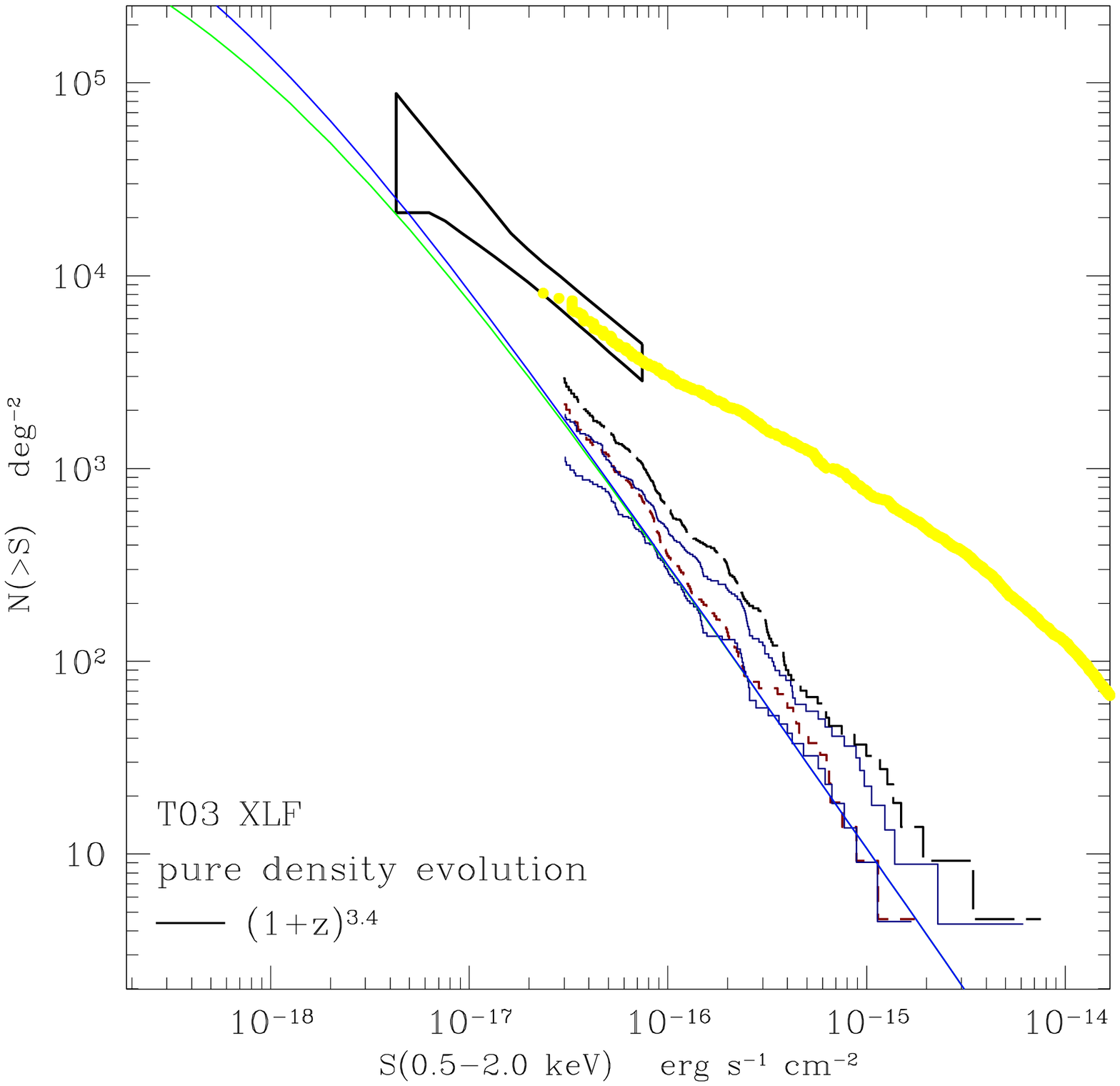}
    \includegraphics[width=0.72\columnwidth]{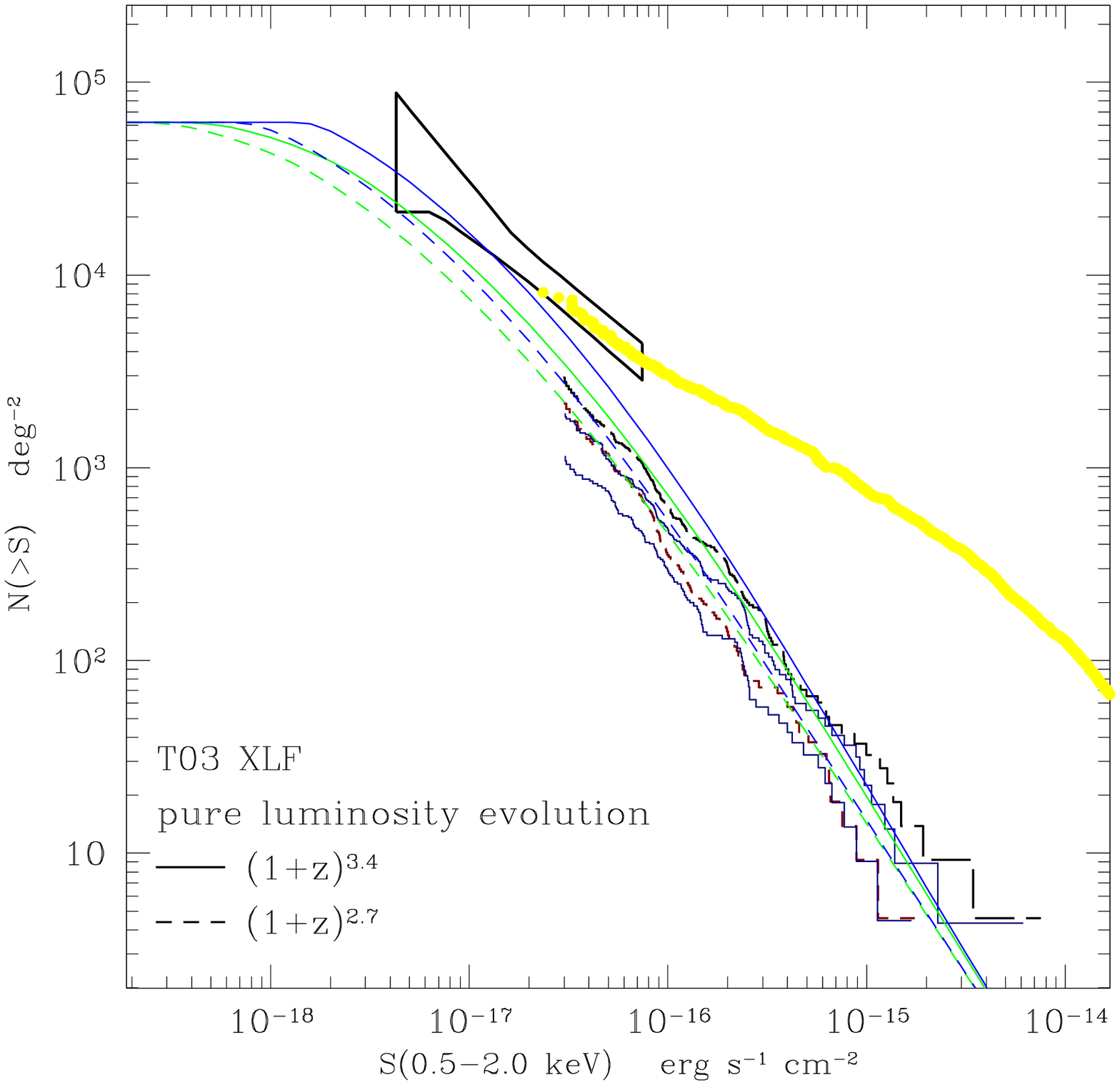}
  }{
    \includegraphics[width=0.497\textwidth]{xcts_takeuden.ps}
    \includegraphics[width=0.497\textwidth]{xcts_takeulum.ps}
  }
  \end{center}
  \caption { \figuraacolori X-ray counts derived from the IRAS
    luminosity function. The continuous and dashed curves, which
    converge at fluxes $\gtrsim 10^{-15}$ \ergscmq, show
    the number counts from the integration of the T03 XLF (upper, blue
    curves: integration and evolution performed up to $z=2$; lower, green
    curves: integration up to $z=2$ with evolution stopped at $z=1$;
    continuous curves: $\etalum=3.4$; dashed curves:
    $\etalum=2.7$). Other symbols as in Fig.~(\ref{fi:xcts-baueretal}).
    {\em Left panel:} source counts from the T03 XLF
    with pure density evolution ($\etaden=3.4$).  {\em Right panel:}
    same XLF, but with pure luminosity evolutions ($\etalum=3.4$,
    continuous curves and $\etalum=2.7$, dashed curves).  
  \label{fi:xcts-takeu} }
\end{figure*}

As a final comment we notice that the scatter in the observed \lognlogs
(both the systematic scatter between counts derived with different
selection criteria and the Poissonian error on the single \lognlogsa,
Fig.~\ref{fi:xcts-baueretal}) does not allow in itself to disentangle
between density or luminosity evolution, but the above results
strengthen the indication in favour of a luminosity evolution with
$\etalum\sim 2.7$ already emerged in the discussion of the XLFs of
\S~\ref{sec:xlfevol}.

We have so far explored the expected behaviour of the \lognlogs by
integrating up to $z=2$. However, it is known that star forming
galaxies do exist and emit X-rays at redshifts $z\sim 3$, at least in
the form of Lyman-break galaxies \citep{nandra02}. Nonetheless, our
parameterisation may still be justified as very little is known of the
cosmic evolution of galaxies beyond $z\sim 1$--2, and it is not
excluded that galaxies become less luminous, which for practical
purposes would not be too far from the case hereby considered.

In order to investigate the number counts from galaxies at larger
redshifts we consider the following scenario: galaxies evolve in
luminosity up to a redshift $z_{\rm ev}$, and evolve no more beyond,
while the integral is performed up to $z=4$.  
By assuming $\etalum=2.7$, we find that the integrated counts are
consistent with the observed ones if $z_{\rm ev}\lesssim 2$.

A similar argument could be made for pure density evolution: in this
case, the observed all-inclusive X-ray counts would not be
outnumbered, but in order not to exceed the cosmic infrared
background, one must have $z_{\rm ev}\lesssim 2$ (see \S~\ref{sec:csfh}).


\section{X-ray counts and the cosmic star formation history}
\label{sec:csfh}

A first attempt in using the evolution of the XLF to put constraints
on the cosmic star formation history (CSFH) was made in \cite{colin},
where the observed X-ray luminosity density was converted to a SFR
density at the redshifts $z=0.27$ and 0.79. Here we investigate a
different possibility, which is to use the X-ray number counts to
constrain the CSFH at higher redshifts ($z\lesssim 4$).

The shape of the CSFH beyond $z\sim 1$--2 is still debated. Results
from UV observations \citep{lilly96,madau96} pointed towards a peak
around $z\sim 2$ with a SFR($z=2$) $\sim 10$ times that at $z=0$.
\citet{pettini98} applied a correction for dust extinction, which
raised the peak SFR by a factor 2 and at the same time shifted the
peak redshift at $z\sim 3$. However, sub-mm observations with the
SCUBA camera could not be reconciled with the UV-derived, extinction
corrected CSFH, leading \citet{blain99a} to propose the existence of a
new population of luminous, strongly obscured, high redshift
($z\lesssim 5$) galaxies. The CSFH derived from sub-mm observations
have a peak at $z\sim 3$ with a SFR $\sim 100$ times the current
($z=0$) one; at higher redshifts the data allow the SFR to stay
constant or decline.

In \S~\ref{sec:counts-integration} we have discussed a particular
case, in which the galaxies are distributed according to the evolved
T03 XLF in the redshift interval 0--4, and we found that the XLF is
not allowed to evolve beyond $z\sim 1$--2 in order not to violate the
observed X-ray source counts. This would be consistent with a SFR that
peaks at $z\sim 1$--2.

To complement our treatment, we follow the approach developed in
\citet{white98}, \citet{ghosh01} and \citet{ptak01}: under the
assumption that the X-ray luminosity of a galaxy results from the
integrated luminosity of its populations of High and Low Mass X-ray
Binaries (HMXB and LMXB), it is possible to derive the evolution of
the X-ray luminosity by considering the galaxy star formation history,
the number ratio of HMXB and LMXB, and the timescales needed to switch
on the X-ray emission.


In order to derive a prediction for the evolution of an X-ray
luminosity function, \citet{ghosh01} and \citet{ptak01} considered
different models of the CSFH in the redshift interval $z=0$--8
representing the different determinations of the CSFH. Here we consider
the two models discussed in \citet{ptak01}:
\begin{itemize}
\item a `peak' model, which presents a rise of the SFR up $z\sim 1.5$
followed by an exponential decline, closely resembling the CSFH
determination by \citet{madau98} (hereafter: `peak-M' model);
\item a `peak+Gaussian' model, which adds a Gaussian component at $z=1.7$
to the `peak' model, in order to account for high redshift data,
namely the sub-mm observations by \citet{blain99a,blain99b}
(hereafter: `peak-G' model). 
\end{itemize}

We derive X-ray number counts by considering the T03 XLF and applying
a pure luminosity evolution from the models in \citet{ptak01}.  The
results are shown in Fig.~(\ref{fi:xcts-takeuptak}). We notice that
our predicted counts are in agreement with those derived in
\citet{ptak01} by converting number counts in the Hubble Deep Field
from the B band to the X-rays, for fluxes in the 0.5--2 keV band $\le
3\e{-17}$ \ergs; altogether, the larger flux range ($\lesssim 10^{-14}$
\ergs) here explored the allows a direct check against the observed
source counts.

\begin{figure}[tp]    
  \begin{center}
  \includegraphics[width=\columnwidth]{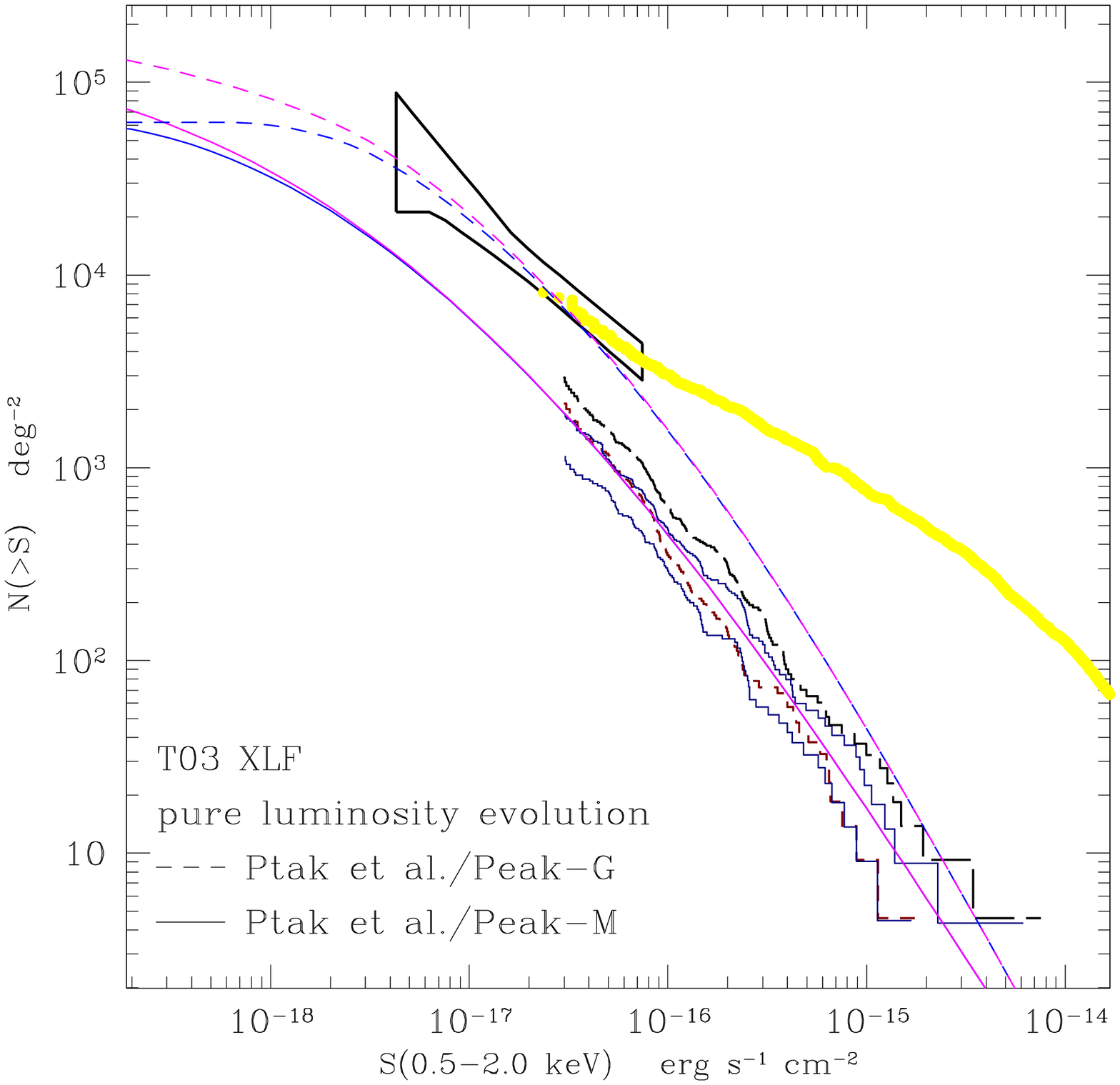}
  \end{center} 
   \caption { \figuraacolori X-ray counts derived from the IRAS
  luminosity function \citep{takeu03,takeu03err} with pure
  luminosity evolution according to \citet{ptak01}. The solid
  curves show the integrated counts assuming the `peak-M' model for the
  cosmic star formation history, with integration and evolution
  performed from $z=0$ to 2 (lower, blue curve) or 4 (upper, pink
  curve). Dashed curves: same as above, for the `peak-G' model.
  Other symbols as in Fig.~(\ref{fi:xcts-baueretal}).
  \label{fi:xcts-takeuptak} }
\end{figure}

We find that the model of the XLF evolution derived from the `peak-G'
model of the CSFH, applied to the T03 XLF, does not agree with the
observed X-ray \lognlogsa. On the other hand, the `peak-M' model is a
good description of the observed source counts and closely resembles
the source counts we have obtained by integrating the T03 XLF up to
$z=2$ with $\etalum=2.7$ and stopping the evolution at $z=1$. Thus it
seems that the CSFH derived from sub-mm observations cannot be easily
reconciled with the X-ray number counts. It might be, for example,
that the sub-mm population of galaxies is obscured in the X-rays,
or that a significant fraction of these galaxies hosts an AGN, hence
the true SFR at high redshift would be smaller. This latter hypothesis
is also reinforced by the fact that the `peak-G model' of the CSFH
leads to a predicted value of the flux of the cosmic IR background
$\sim 3$ times larger than the current upper limit, as shown below.
However, further consideration of this issue would be beyond the
scope of this paper.


Additional constraints on the evolution properties here examined come
from the consideration that total flux of the galaxies should not
exceed the cosmic background levels. The most stringent one comes from
the Cosmic Infrared Background (CIB; see \citealt{hauser01} for a
review) since it is believed that star forming galaxies make the bulk
of the CIB, while they only make around 50\% of the radio background
\citep{haarsma98}, or a few percent of the X-ray background (RCS03).
Recent results have shown that AGN contribute $\sim 2\%$ of the CIB at
$60\mu$ \citep{silva04}.

Due to the extreme difficulty of the $60\mu$ CIB level measurement,
its actual flux value is still debated. A tentative detection
\citep{finkbeiner00} fixed the flux value at $\nu I_\nu=28\pm 7$
\nWmqsr, while two upper limits have been set at $12$ and $10$ \nWmqsr
(\citealt{kashlinsky00} and \citealt{dwek01}, respectively).
\citet{miville02} also fixed an upper limit at 13.5 \nWmqsr, with a
tentative detection at 9 \nWmqsr.  While we refer the reader to
\citet{hauser01} for a complete discussion of the CIB determination,
in the following we will assume as the CIB flux $10$ \nWmqsr,
corresponding to the most stringent upper limit. Thus, the following
estimates of the contribution to the CIB should be regarded as lower
limits.

The cosmic background flux from the integration of known sources may
be defined as
\begin{equation}
  \label{eq:fondo1}
  F = \int_{-\infty}^{+\infty} \de\Log S\, \frac{\de N}{\de\Log S}\, S
\end{equation}
of the differential \lognlogsa, which in turn is related to the LF by
\begin{equation}
  \label{eq:fondo2}
\frac{\de N}{\de\Log S} = \int_{z_{\rm min}}^{z_{\rm max}} \de z 
\ \varphi(\Log L(S,z)) \, K(z)
\, \frac{\de V}{\de z}
\end{equation}
where $K(z)$ is the K-correction, here calculated after the spectral
model of M82 from the GRASIL code \citep{grasil}.

Taking the T03 LF as a reference and assuming luminosity evolution up
to $z=4$, we found the contribution levels to the $60\mu$ CIB to be in
the range $60\%$--$90\%$, depending on the details of evolution: the
lower figure was obtained by considering $\etalum=2.7$ and evolution
up to $z=2$, the upper one by considering $\etalum=3.4$ and evolution
up to $z=4$. Had we considered density evolution, the contribution
range would have been $60\%$--$110\%$ (lower figure: $\etaden=2.7$,
$z_{\rm ev}=2$; upper figure: $\etaden=3.4$, $z_{\rm ev}=4$), thus
ruling out density evolution beyond $z_{\rm ev}\sim 2$.

The main conclusion that may be drawn, is that the galaxy LF may not
probably sustain an evolution as strong as $\etaden=3.4$ or
$\etalum=3.4$ beyond $z\sim 2$. Milder evolution like $\etalum=2.7$
might be sustained up to larger redshifts, with the {\em caveat} that
future analysis may lower the flux level of the $60\mu$ CIB.



In a complementary way, by using the `peak-M' and `peak-G' models
introduced above as prescriptions for the evolution of the infrared
luminosity of galaxies, and consistently with the results from the
\lognlogs analysis, we find that the the contribution to the CIB by
sources distributed according to the T03 LF with the `peak-M' model
for evolution is about $60\%$, while the `peak-G' model is ruled out
as the predicted flux from integrated sources is $\sim 290\%$ of the
CIB (due to the bell-like shape of these models, these percentages do
not depend significantly on the maximum redshifts of integration and
evolution).


\section{X-ray number counts and deep radio surveys}
\label{sec:radiocts}

Although not directly related to the determination of the XLF
properties discussed in the preceding Sections, it is of interest to
consider the X-ray number counts that can be derived from the radio
sub-mJy population, which is commonly believed to include high
redshift star forming galaxies as a major component.  A prediction of
the X-ray number counts based on the radio \lognlogs for the sub-mJy
population was already derived in RCS03. However, it was not possible
in RCS03 to compare the predicted counts with observed ones, since the
latter were not available at that time.  

\begin{figure}[tp]    
  \begin{center}
  \includegraphics[width=\columnwidth]{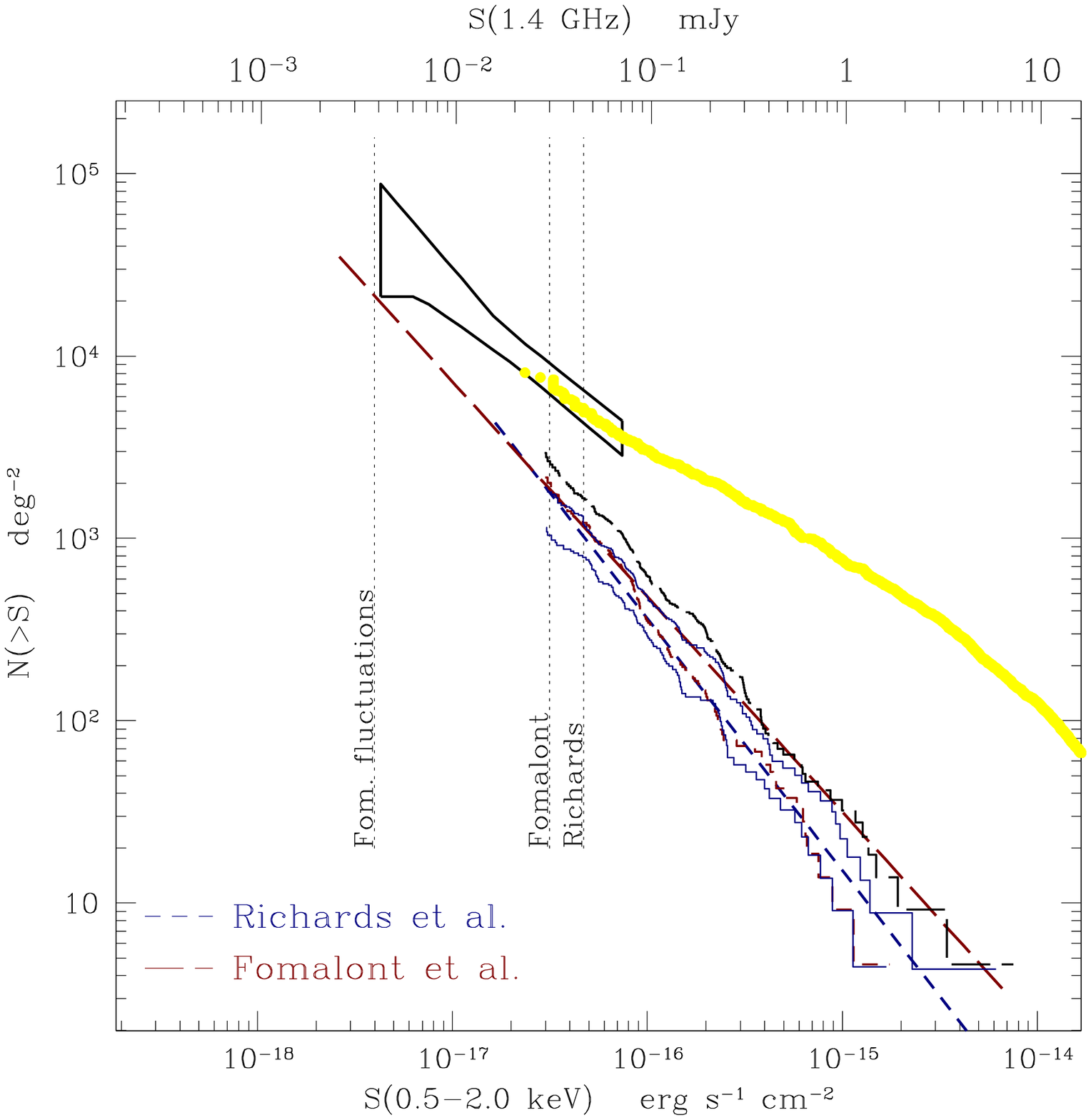}
  \end{center} 
  \caption{ \figuraacolori X-ray counts derived from deep radio
    \lognlogs in the 0.5--2.0 keV band (see RCS03).  The short-dashed
    (blue) lines represent the 1.4 GHz \lognlogs \citep{vla00}.  The
    long-dashed (red) lines represent the \lognlogs by
    \citet{fomalont91}. The vertical dotted lines show the limiting
    sensitivities for the radio surveys. For easier reading, the
    horizontal scale is shown both in radio (upper) and X-ray (lower)
    fluxes, assuming the conversion factors of RCS03.  Other symbols
    as in Fig.~(\ref{fi:xcts-baueretal}). }
  \label{fig_lognlogs} 
\end{figure}

The radio \lognlogs converted to the X-rays are shown in
Fig.~(\ref{fig_lognlogs}) as the short- \citep{vla00} and long-dashed
\citep{fomalont91} lines. The agreement with the observed counts is
very good, with the \citeauthor{vla00} \lognlogs closely following the
counts selected with the X-ray/optical flux ratio, and with the
\citeauthor{fomalont91} \lognlogs resembling the counts selected with
the Bayesian criterion by \citet{colin}.  Since \citet{fomalont91}
checked that the source density at 4 $\mu$Jy (5 GHz) derived from
fluctuation analysis was consistent with a simple extrapolation of the
\lognlogsa, we extended the corresponding X-ray counts in down to the
flux level of the radio background fluctuations.  This simple
extrapolation suggests that galaxies may become more numerous than AGN
at some flux below $\sim 10^{-17}$ \ergscmq.

From Fig.~(\ref{fig_lognlogs}) it is seen that X-ray
sources with fluxes below $\sim 10^{-15}$ \ergscmq\ should have a
radio flux below $\sim 1$ mJy. This may be checked as follows: we
selected, from the sources with $\fxott <-1$ previously discussed, the
subsample lying in the CDFN area (97 sources); this subsample was
cross-correlated with the 1.4GHz catalogue of the same area
\citep{vla00}. We found that all the X-ray sources with a radio
counterpart have a 1.4GHz flux below 2 mJy, with most of them around
0.1 mJy.  It has been proposed that a major fraction of the radio
sources building the `sub-mJy population', which accounts for the
majority of the radio number counts below $\sim 0.5$ mJy
\citep{windhorst85} and about half of the radio cosmic background at
1.4GHz \citep{haarsma00}, are star forming galaxies at redshifts
$0.5\lesssim z\lesssim 1$ \citep{windhorst90}. Thus our findings
contribute to reinforce this view.

\section{Conclusions}
\label{sec:counts-finalremarks}

By adopting well known relationships linking the IR, radio, optical 
and X-ray emissions of normal spiral and starburst galaxies we have 
transformed available LFs of these objects from samples selected at
IR, radio and optical wavelengths into corresponding X-ray luminosity 
functions (XLF). From the comparison of these XLFs among themselves
and with the X-ray selected XLFs of \citet{colin} it is 
concluded that:

\begin{itemize}
\item A clear prediction for a local ($z=0$) XLF emerges from the
  comparison of the infrared, radio and optical LFs
  (Fig.~\ref{fi:xraytake}): the derived XLFs agree within a factor 2
  in the luminosity interval $10^{40}$--$10^{41}$ \ergs\ encompassing
  the knee region after which all XLFs steepen toward higher
  luminosities; although departures at lower and higher luminosities
  are present, the average local X-ray luminosity density ($\sim
  (3\e{37} \pm 30\%)$ \ergs\ Mpc$^{-1}$) appears to be well defined
  (Table~\ref{tab:luminositydensity}).

\item There is a remarkably good agreement between the two best
  defined XLFs from the IRAS sample (T03) and W03, both based on the
  largest galaxy samples with $\sim 10^4$ objects each and on very
  different selection criteria: the W03 XLFs are all well reproduced
  by the local T03 XLF once a luminosity evolution with coefficient
  $\etalum \sim 3$ is applied. Thus the
  T03 XLF may be considered representative of the galaxies XLF up to
  $z\sim 1$.

\item None of the XLFs derived from the infrared and radio local
  samples ($z<0.3$) can match the higher luminosities ($\LX>3\e{41}$,
  Figs.~\ref{fi:xlf-normanetal}a,c) of the observed \citep{colin} XLFs
  with their quoted (Table~\ref{tab:LFs}) density evolution parameters
  ($\etaden$).  However, a very good match is achieved, particularly
  in the high redshift bin, by adopting pure luminosity evolutions
  with parameters $\etalum$ approximately equal to $\etaden$ of the
  respective samples (Figs.~\ref{fi:xlf-normanetal}b,d).  Among the
  XLFs from the radio, IR and optical samples selected over a wider
  redshift range (P04, C89/H00, W03) only those from the blue sample
  (W03) match the observed XLFs in both redshift bins and in the
  luminosity intervals in which they are defined.  
\end{itemize}

By combining the above findings we conclude that the local XLF
derived from IRAS T03 sample provides a consistent description of the
local XLF of star forming galaxies and that pure luminosity evolution
of the form $(1+z)^\etalum$, with $\etalum\sim 3$ (consistently with
$\etalum\sim 2.7$ as found by \citealt{colin}, and within errors of
the $\etaden=3.4\pm 0.7$ coefficient given for density evolution by
T03), is favoured over density evolution providing a good
representation of available data as derived from radio, optical and
X-ray samples. We note, however, that the excess of the observed XLF
at higher luminosities in the low redshift bin may be caused by a
slight pollution by AGN in the \citet{colin} sample.

By comparing the number counts from our adopted XLF from the T03
sample with those derived from observed X-ray samples of star forming
galaxies, and with the total X-ray number counts, we reach the
following conclusions:

\begin{itemize}
\item The integral number counts from samples of star forming galaxies
show very similar slopes but their normalisations are dispersed within
a factor $\sim 2$. The observed fraction of star forming galaxies at
the flux limit of the \chandra\ Deep Field surveys ($\sim 3\e{-17}$
\ergscmq) is $(30\pm 12)\%$.

\item The counts obtained by adopting different values of the
  luminosity evolution parameter $\etalum$ and redshift cut-offs in
  the evolution indicate that a good representation of the observed
  data is obtained with $\etalum\sim 2.7$ and a cut-off of the
  evolution in the redshift interval 1-2, which is believed to host
  the peak of the star formation rate. Much stronger evolution, say
  $\etalum = 3.4$ as adopted in our analysis, is not consistent with
  the observed counts unless one stops the evolution of the XLF at a
  redshift $z\sim 1$. 
\item Additional, supportive evidence of the above results is provided
  by the very good agreement of the X-ray integral counts derived from
  the radio \lognlogs of the sub-mJy population, believed to be mainly
  representative of the star forming galaxies, with the observed
  counts and those from the integration of the T03 XLF (compare
  Fig.~\ref{fi:xcts-baueretal} and Fig.~\ref{fig_lognlogs}).
\end{itemize}

We have complemented the analysis of the \lognlogs by adopting our
proposed XLF in the framework of the cosmic star formation history
(CSFH) as parameterised in the models of \citet{ptak01}, finding that
the observed counts are only consistent with model `peak-M' of these
authors, i.e.\ with a model of the CSFH resembling the data from
\citet{madau98}. We note that the CSFH model suggested by sub-mm
observations \citep{blain99a} appears to be ruled out both by the
observed X-ray number counts and by the flux level of the Cosmic
Infrared Background (CIB).

As a further check on our results, we have estimated the contribution
of the galaxies in the T03 sample to the CIB. Pure luminosity
evolution with $\etalum\sim 2.7$ is consistent with the current CIB
upper limit; moreover, the X-ray counts predicted by the `peak-M'
model are almost coincident with those obtained from the integration
the T03 XLF with $\etalum=2.7$.  We have also found that the evolution
implied by the `peak-G' model on the T03 LF would exceed the CIB limit
by a factor $\sim 3$.


It may be predicted that $6\e{3}$--$10^{4}$ sources deg$^{-2}$ would
be present with fluxes $\geq 10^{-17}$ \ergscmq. This figure may be
compared with the AGN number density between $4\e{3}$ and $10^4$
deg$^{-2}$ at fluxes below $\sim 10^{-16}$ \ergscmq, as predicted by
AGN models \citep{comastri95,ueda03}. It follows that, if we
conservatively assume the upper estimate, the X-ray number counts of
normal galaxies should overcome the counts from AGN at fluxes
$\lesssim 10^{-17}$ \ergscmq\ ($6\e{-18}$ \ergscmq\ in the case of the
W03 XLF or the T03 XLF with the `peak-M' evolution).

The \chandra\ Deep Field surveys have provided for the first time a
sizable number (100--200, depending on selection criteria) of X-ray
detections of high redshift galaxies.  However, further studies on
galaxy evolution would greatly benefit from enlarged and/or deeper
samples. If a 0.5--2 keV flux of $10^{-17}$ \ergscmq\ ($\sim 3$ times
fainter than the current flux limit of the \chandra\ Deep Field North)
could be reached, it would be possible to detect galaxies with
$\LX=10^{41}$ \ergs\ placed at $z=1.3$, or with $\LX=3\e{41}$ \ergs\
at $z=2$. Our models show that the density of galaxies within this
luminosity and redshift ranges should be $\sim 4000$ deg$^{-2}$; that
is considering over an area of, say, 100 arcmin$^2$, $\sim 100$ could
be detected.  In principle it would then be possible to derive the
high luminosity tail of the galaxies XLF in the
$z=1.3$--2 redshift bin, thus placing a significant constraint on the
evolution of the SFR.


Although these studies are feasible in principle with \chandra, they
would require a huge investment of time. From our models it is also
apparent that more stringent constraints could be achieved by pushing
the detection limit toward $10^{-18}$, so that XEUS would be more
suited for this purpose, if the goal of a PSF with diameter $\lesssim
2\arcsec$ can be achieved (given the predicted number of galaxies
around $10^{-18}$ \ergscmq, a significantly larger PSF would lead to
source confusion).

\begin{acknowledgements}
  We thank Francesca Pozzi for many useful discussions, and Ann
  Hornschemeier and Tsutomu T. Takeuchi and for their kindness in
  answering our many questions.  Finally, we thank the anonymous
  referee for extremely useful suggestions leading to an improved
  version of this work. This work was supported by the MIUR grants
  COFIN-03-02-23 and by the INAF grant PRIN/270.
\end{acknowledgements}


\begin{thebibliography}{}


\bibitem[\protect\astroncite{Alexander et al.}{2003}]{alexander03}Alexander D.~M., Bauer F.~E., Brandt W.~N., et al.\ 2003, AJ,  126, 539 


\bibitem[Avni \& Tananbaum(1986)]{avnitan86}Avni, Y.~\& 
Tananbaum, H.\ 1986, \apj, 305, 83 


\bibitem[\protect\astroncite{Bauer et~al.}{2002}]{bauer02}Bauer, F.E., Alexander, D.M., Brandt, W.N. et al.~2002, AJ, 124, 2351

\bibitem[\protect\astroncite{Bauer et~al.}{2004}]{bauer04}Bauer, F.E., Alexander, D.M., Brandt, W.N. et al.~2004, AJ, 128, 2048


\bibitem[\protect\astroncite{Blain et al.}{1999b}]{blain99b}Blain A.~W., Jameson A., Smail I., et al.\ 1999, MNRAS,  309, 715 

\bibitem[\protect\astroncite{Blain et al.}{1999a}]{blain99a}Blain A.~W., Smail I., Ivison R.~J., Kneib J.-P.\ 1999, MNRAS,  302, 632 

\bibitem[\protect\astroncite{Comastri et~al.}{1995}]{comastri95}Comastri A., Setti G., Zamorani G. \& Hasinger G. 1995, \aap\ 296, 1


\bibitem[\protect\astroncite{Condon}{1989}]{condon89}Condon, J. J. 1989, \apj\ 338, 13

\bibitem[\protect\astroncite{Condon}{1992}]{cond92}Condon J.J. 1992, ARA\&A 30, 575

\bibitem[\protect\astroncite{David et~al.}{1992}]{djf92}David L.P., Jones C. \& Forman W. 1992, ApJ 388, 82



\bibitem[\protect\astroncite{de Jong et al.}{1984}]{dejong84}de Jong T., Clegg P.~E., Rowan-Robinson M., et al.\ 1984, ApJ,  278, L67 

\bibitem[\protect\astroncite{de Vaucouleurs et 
al.}{1991}]{rc3}de Vaucouleurs G., de Vaucouleurs A., 
Corwin H.~G., et al.\ 1991, Third Reference Catalogue of Bright
Galaxies, Springer-Verlag, Berlin Heidelberg New York (RC3)


\bibitem[\protect\astroncite{Dwek}{2001}]{dwek01}Dwek E. 2001, proc.\ symp.\ "The Extragalactic Infrared Background and
its Cosmological Implications", IAU Symp.\ 204, (eds.\ Harwit M. \&
Hauser M. G.), {\em astroph/0105363}
 

\bibitem[\protect\astroncite{Elbaz et al.}{2002}]{elbaz02}Elbaz D., Cesarsky C.~J., Chanial P., et al.\ 2002, A\&A,  384, 848 


\bibitem[\protect\astroncite{Fabbiano}{1989}]{fabbiano89}Fabbiano G. 1989, ARA\&A 27, 87


\bibitem[\protect\astroncite{Fabbiano et al.}{1988}]{fgt88}Fabbiano G., Gioia I.M. \& Trinchieri G. 1988, ApJ 324, 749

\bibitem[\protect\astroncite{Fabbiano \& Shapley}{2002}]{fabshap02}Fabbiano G., Shapley A.\ 2002, ApJ, 565, 908 

\bibitem[\protect\astroncite{Finkbeiner et al.}{2000}]{finkbeiner00}Finkbeiner D.~P., Davis M., Schlegel D.~J.\ 2000, ApJ,  544, 81 

\bibitem[\protect\astroncite{Fomalont et al.}{1991}]{fomalont91}Fomalont E.B., Windhorst R.A., Kristian J.A. \& Kellerman K.I. 1991, AJ 102, 1258

\bibitem[\protect\astroncite{Franceschini et 
al.}{2001}]{franceschini01}Franceschini A., Aussel H., Cesarsky 
C.~J., Elbaz D., Fadda D.\ 2001, A\&A,  378, 1 

\bibitem[\protect\astroncite{Gehrels}{1986}]{gehrels86}Gehrels 
N.\ 1986, ApJ,  303, 336 


\bibitem[\protect\astroncite{Georgakakis et 
al.}{2004}]{georgakakis04}Georgakakis A.~E., Georgantopoulos I., 
Basilakos S., Plionis M., Kolokotronis V.\ 2004, MNRAS,  354, 123 

\bibitem[\protect\astroncite{Georgantopoulos et al.}{1999}]{ioannis99}Georgantopoulos, I., Basilakos, S., \& Plionis, M. 1999, \mnras, 305, L31

\bibitem[\protect\astroncite{Ghosh \& White}{2001}]{ghosh01}Ghosh, P.~\& White, N.~E.\ 2001, \apjl, 559, L97 

\bibitem[\protect\astroncite{Giacconi \& Zamorani}{1987}]{giacconi87}Giacconi R., Zamorani G.\ 1987, ApJ,  313, 20 

\bibitem[\protect\astroncite{Giacconi et al.}{2002}]{giacconi02}Giacconi R., Zirm A., Wang J., et al.\ 2002, ApJS,  139, 369 


\bibitem[\protect\astroncite{Haarsma et al.}{2000}]{haarsma00}Haarsma D.B., Partridge R.B., Windhorst R.A. \& Richards E.A. 2000, ApJ 544, 641
\bibitem[\protect\astroncite{Haarsma \& 
Partridge}{1998}]{haarsma98}Haarsma D.~B., Partridge R.~B.\ 
1998, ApJ,  503, L5 

\bibitem[\protect\astroncite{Hauser \& Dwek}{2001}]{hauser01}Hauser M.~G., Dwek E.\ 2001, ARA\&A,  39, 249 

\bibitem[\protect\astroncite{Hornschemeier et 
al.}{2003}]{hornsch03}Hornschemeier A.~E., Bauer F.~E., 
Alexander D.~M., et al.\ 2003, AJ,  126, 575 

\bibitem[\protect\astroncite{Isobe et al.}{1990}]{isobe90}Isobe T., Feigelson E.~D., Akritas M.~G., Babu G.~J.\ 1990, ApJ,  364, 104 

\bibitem[\protect\astroncite{Kashlinsky \& 
Odenwald}{2000}]{kashlinsky00}Kashlinsky A., Odenwald S.\ 2000, 
ApJ,  528, 74 

\bibitem[\protect\astroncite{Kennicutt}{1998}]{kenn98}Kennicutt R.C. Jr.~1998, ApJ 498, 541


\bibitem[\protect\astroncite{Kinney et al.}{1996}]{kinney96}Kinney A.~L., Calzetti D., Bohlin R.~C., et al.\ 1996, ApJ,  467, 38 

\bibitem[\protect\astroncite{La Valley et al.}{1992}]{asurv}La Valley, M. P., Isobe, T. \& Feigelson, E. D. 1992, BAAS,  24, 839 


\bibitem[\protect\astroncite{Lilly et al.}{1996}]{lilly96}Lilly S.~J., Le Fevre O., Hammer F., Crampton D.\ 1996, ApJ,  460, L1 


\bibitem[\protect\astroncite{Maccacaro et al.}{1988}]{maccacaro88}Maccacaro T., Gioia I.~M., Wolter A., Zamorani G., Stocke J.~T., 1988, ApJ, 
326, 680 


\bibitem[\protect\astroncite{Machalski \& Condon}{1999}]{machalski99}Machalski, J., \& Condon, J. J. 1999, \apjs\ 123, 41


\bibitem[\protect\astroncite{Machalski \& Godlowski}{2000}]{machalski00}Machalski, J., \& Godlowski, W. 2000, \aap\ 360, 463


\bibitem[\protect\astroncite{Madau et al.}{1996}]{madau96}Madau P., Ferguson H.~C., Dickinson M.~E., et al.\ 1996, MNRAS,  283, 1388 


\bibitem[\protect\astroncite{Madau et al.}{1998}]{madau98}Madau P., Pozzetti L., Dickinson M.\ 1998, ApJ,  498, 106 

\bibitem[\protect\astroncite{Madgwick et al.}{2002}]{madgwick02}Madgwick D.~S., Lahav O., Baldry I.~K., et al.\ 2002, MNRAS,  333, 133 

\bibitem[\protect\astroncite{Mazzei et al.}{2001}]{mazzei01}Mazzei P., Aussel H., Xu C., et al.\ 2001, NewA,  6, 265 

\bibitem[\protect\astroncite{Miville-Desch\^enes et al.}{2002}]{miville02}Miville-Desch\^enes M.-A., Lagache G., Puget J.-L.\ 2002, A\&A,  393, 749 

\bibitem[\protect\astroncite{Miyaji \& Griffiths}{2002a}]{miyaji02a}Miyaji T. \& Griffiths R.E. 2002a, ApJ 564, L5


\bibitem[\protect\astroncite{Miyaji \& Griffiths}{2002b}]{miyaji02b}Miyaji T. \& Griffiths R.E. 2002b, Proc. Symp.~``New Visions of
the X-ray Universe in the XMM-Newton and Chandra Era'', ESTEC 2001
{\em astro-ph/0202048}


\bibitem[\protect\astroncite{Moretti et al.}{2003}]{moretti03}Moretti A., Campana S., Lazzati D., Tagliaferri G.\ 2003, ApJ,  588, 696 


\bibitem[\protect\astroncite{Nandra et al.}{2002}]{nandra02}Nandra K., Mushotzky R.F., Arnaud K. et~al.~2002, ApJ 576, 625


\bibitem[\protect\astroncite{Nilson}{1973}]{ugc}Nilson P.\ 1973, Acta Universitatis Upsaliensis. Nova Acta Regiae
  Societatis Scientiarum Upsaliensis - Uppsala Astronomiska
  Observatoriums Annaler, Uppsala: Astronomiska Observatorium.

\bibitem[\protect\astroncite{Norman et~al.}{2004}]{colin}Norman, C., Ptak, A., Hornschemeier, A., Hasinger, G. et al.,
ApJ 607, 721

\bibitem[\protect\astroncite{Page \& Carrera}{2000}]{pageca00}Page, M. J. \& Carrera, F. J.  2000, MNRAS 311, 433


\bibitem[\protect\astroncite{Pettini et al.}{1998}]{pettini98}Pettini M., Kellogg M., Steidel C.~C., et al.\ 1998, ApJ,  508, 539 

\bibitem[\protect\astroncite{Pozzi et al.}{2004}]{pozzi04}Pozzi, F., Gruppioni, C., Oliver, S., Matute, I. et al., ApJ, 609, 122

\bibitem[\protect\astroncite{Ptak et~al.}{2001}]{ptak01}Ptak A., Griffiths R., White N. \& Ghosh P. 2001, ApJ 559, L91


\bibitem[\protect\astroncite{Ranalli et al.}{2003}]{rcs03}Ranalli P., Comastri A., Setti G.\ 2003, A\&A,  399, 39 

\bibitem[\protect\astroncite{Revnivtsev et al.}{2005}]{revnivtsev05}
Revnivtsev M., Gilfanov M., Jahoda K., Sunyaev R.\ 2005, submitted to
\aap, {\em astro-ph/0412304}


\bibitem[\protect\astroncite{Richards}{2000}]{vla00}Richards E.A. 2000, ApJ 533, 611


\bibitem[\protect\astroncite{Rush et~al.}{1993}]{rush}Rush B., Malkan M. \& Spinoglio L. 1993, ApJS 89, 1

\bibitem[\protect\astroncite{Sadler et al.}{2002}]{sadler02}Sadler E.~M., Jackson C.~A., Cannon R.~D., et al.\ 2002, MNRAS,  329, 227 

\bibitem[\protect\astroncite{Sandage \& Tammann}{1981}]{RSA}Sandage A.R. \& Tammann G.A. 1981, A Revised Shapley-Ames Catalog of Bright Galaxies (RSA),
Washington: Carnegie Institution of Washington


\bibitem[\protect\astroncite{Saunders et al.}{1990}]{saun90}Saunders, W., Rowan-Robinson, M., Lawrence, A. et al.~1990, \mnras, 242, 318


\bibitem[\protect\astroncite{Saunders et al.}{2000}]{saun00}Saunders W., Sutherland W.~J., Maddox S.~J., et al.\ 2000, MNRAS,  317, 55 

\bibitem[\protect\astroncite{Schmidt et al.}{1996}]{schmidt1996}Schmidt K.-H., Boller T., Voges W.\ 1996, Proc. `R\"ontgenstrahlung
from the Universe', eds. Zimmermann, H.U.; Trümper, J.; and Yorke,
H.; MPE Report 263, 395


\bibitem[\protect\astroncite{Schmidt}{1968}]{schmidt68}Schmidt, M. 1968, ApJ 151, 393

\bibitem[\protect\astroncite{Serjeant et al.}{2004}]{serje04}Serjeant, S., Carrami\~nana, A., Gonz\'ales-Solares, E. et al.~2004,
\mnras, 355, 813


\bibitem[\protect\astroncite{Serjeant et al.}{2001}]{serje01}Serjeant, S., Eftathiou, A., Oliver, S. et al.~2001, \mnras, 322, 262


\bibitem[\protect\astroncite{Shapley et al.}{2001}]{shapley01}Shapley A., Fabbiano G., Eskridge P.~B.\ 2001, ApJS,  137, 139 

\bibitem[\protect\astroncite{Silva et al.}{1998}]{grasil}Silva L., Granato G.~L., Bressan A., Danese L.\ 1998, ApJ,  509, 103 

\bibitem[\protect\astroncite{Silva et al.}{2004}]{silva04}Silva L., Maiolino R., Granato G.~L.\ 2004, MNRAS, 355, 973 


\bibitem[\protect\astroncite{Tajer et al.}{2005}]{tajer05}Tajer M., Trinchieri G., Wolter A., et al.\ 2005, accepted by \aap,
 {\em astro-ph/0412588}

\bibitem[\protect\astroncite{Takeuchi et al.}{2003}]{takeu03}Takeuchi, T. T., Yoshikawa, K., \& Ishii, T. T. 2003, \apj, 587, L89

\bibitem[\protect\astroncite{Takeuchi et al.}{2004}]{takeu03err}Takeuchi T.~T., Yoshikawa K., Ishii T.~T.\ 2004, ApJ,  606, L171 

\bibitem[\protect\astroncite{Ueda et al.}{2003}]{ueda03}Ueda 
Y., Akiyama M., Ohta K., Miyaji T.\ 2003, ApJ,  598, 886 

\bibitem[\protect\astroncite{White \& Ghosh}{1998}]{white98}White, N.~E.~\& Ghosh, P.\ 1998, \apjl, 504, L31 

\bibitem[\protect\astroncite{Windhorst et al.}{1990}]{windhorst90}Windhorst R.A., Mathis D.F. \& Neuschaefer L.W. 1990, Proc.~Symp.~%
``Evolution of the universe of galaxies: Edwin Hubble Centennial Symposium'',
ASP Conference Series 10, 389


\bibitem[\protect\astroncite{Windhorst et al.}{1985}]{windhorst85}Windhorst R.A., Miley G.K., Owen F.N., Kron R.G. \& Koo D.C. 1985, ApJ 289, 494

\bibitem[\protect\astroncite{Wolf et al.}{2003}]{wolf03}Wolf C., Meisenheimer K., Rix H.-W., et al.\ 2003, A\&A,  401, 73 


\end{thebibliography}
\end{document}